\documentclass[fleqn,usenatbib]{aa}  

\usepackage{graphicx}
\usepackage{txfonts}
\usepackage{hyperref}
\hypersetup{
    colorlinks=true,
    citecolor=blue,
    linkcolor=blue,
    filecolor=blue,      
    urlcolor=blue,
    }
\usepackage{graphicx}	
\usepackage{subcaption}
\usepackage{amsmath}	
\usepackage{cleveref}
\usepackage{xspace}
\usepackage{ulem}

\newcommand{\angstrom}{\mbox{\normalfont\AA}\xspace}
\newcommand{\ropt}{$\rm R_{\textup{opt}}$\xspace}
\newcommand{\solarmass}{\rm M_{\odot}}
\newcommand{\km}{km s$^{-1}$\xspace}
\newcommand{\camila}{}

\begin{document}

   \title{First resolved stellar halo kinematics of a MW-mass galaxy outside the Local Group: A flat counter-rotating halo in NGC 4945}


   \author{Camila Beltrand
          \inst{1},
          Antonela Monachesi\inst{1}, 
          Richard D'Souza\inst{2},
          Eric F. Bell,\inst{3},
Roelof S. de Jong\inst{4},
Facundo A. Gomez,\inst{1}
Jeremy Bailin\inst{5},
In Sung Jang\inst{6}
\and
Adam Smercina\inst{7}
          }
   \institute{Departamento de Astronom\'ia, Universidad de La Serena, Av. Ra\'ul Bitr\'an 1305, La Serena, Chile\\
              \email{camila.beltrand@userena.cl}     
              \and
              Vatican Observatory, Specola Vaticana, V-00120, Vatican City State
              \and
              Department of Astronomy, University of Michigan, 311 West Hall, 1085 South University Ave., Ann Arbor, MI 48109-1107, USA
              \and
              Leibniz-Institut f\"{u}r Astrophysik Potsdam (AIP), An der Sternwarte 16, 14482 Potsdam, Germany
              \and
              Department of Physics and Astronomy, University of Alabama, Box 870324, Tuscaloosa, AL 35487-0324, USA
              \and
              Department of Astronomy \& Astrophysics, University of Chicago, 5640 S.\ Ellis Avenue, Chicago, IL 60637, USA 
              \and
              Department of Astronomy, University of Washington, Box 351580, Seattle, WA 98195-1580, USA
             }

   \date{Received ; accepted}

 
  \abstract
   {Stellar halos of galaxies, primarily formed through the accretion and merger of smaller objects, are an important tool to understand the hierarchical mass assembly of galaxies. However, the inner regions of stellar halos in disk galaxies are predicted to have an in-situ component that is expected to be prominent along the major axis. Kinematic information is crucial to disentangle the contribution of the in-situ component from the accreted stellar halos. The low surface brightness of stellar halos makes it inaccessible with traditional integrated light spectroscopy. In this work, using a novel technique, we study the kinematics of the stellar halo of the edge-on galaxy NGC 4945. We couple new deep Multi Unit Spectroscopic Explorer spectroscopic observations with existing Hubble Space Telescope imaging data to spectroscopically measure the line-of-sight (LOS) heliocentric velocity and velocity dispersion in two fields at a galactocentric distance of 12.2 kpc (outer disk field) and 34.6 kpc (stellar halo field) along NGC 4945 major axis, by stacking individual spectra of red giant branch and asymptotic giant branch stars. We obtain a LOS velocity and dispersion of 673 $\pm$ 11 \km and 73 $\pm$ 14 \km, respectively, for the outer disk field. This is consistent with the mean HI velocity of the disk at that distance. For the halo field we obtain a LOS velocity and dispersion of 519 $\pm$ 12 \km and 42 $\pm$ 22 \km. The halo fields' velocity measurement is within $\sim$40 \km from the systemic LOS velocity of NGC 4945, which is 563 \km, suggesting that its stellar halo at 34.6 kpc along the major axis is counter-rotating and is of likely accretion origin. This provides the first ever kinematic measurement of the stellar halo of a Milky Way-mass galaxy outside the Local Group from its resolved stellar population, and establishes a powerful technique for measuring the velocity field of the stellar halos of nearby galaxies.}

\titlerunning{Resolved stellar halo kinematics of NGC 4945}
\authorrunning{C. Beltrand et al.}
   \maketitle
%

\section{Introduction} \label{intro}

Stellar halos emerge as a direct consequence of the hierarchical processes governing galaxy formation. Within the framework of the $\Lambda$CDM (lambda cold dark matter) cosmological model, the accretion and tidal disruption of multiple satellite galaxies \citep{searle1978} by Milk Way-like galaxies creates a low-surface brightness stellar halo around it \citep{bullock2005,cooper2010}. This faint component \citep[$\mu_V \ge 28$,][]{bullock2005} contains only a small fraction of the main galaxy's light (only 1-10 percent), and extends to approximately 100-200 kpc around the galaxy. Recent accretions often leave a large amount of substructure in their outskirts in the form of stellar streams as well as small satellite galaxies \citep{bullock2005,bell2008,johnston2008,cooper2010,vera2022}.

Theory and observations have demonstrated that the variations found in the stellar halos of nearby Milky Way (MW) mass galaxies showcase the diversity of their assembly histories \citep{cooper2010,gomez2012,harmsen2017,  monachesi2019}. These halos primarily reflect the properties of their most massive satellites that were previously accreted \citep{dsouza2018MNRAS,monachesi2019,smercina2020}, which form the majority constituent of what is commonly referred to as the accreted component of the stellar halo.

In addition to a dominant accreted component, theoretical works predict that a small mass fraction of stellar halos comes from stars that were not accreted but instead formed originally within the main potential well of the galaxy \citep{abadi2006,zolotov2009,cooper2015}. This in-situ halo should be more metal-rich than the accreted halo, as its stars were formed in the main, more massive galaxy \citep{monachesi2019} and it is predicted to be more prominent along the major axis of the galactic disk \citep{pillepich2015,monachesi2016b}. The presence of these in-situ halo stars interferes with the characterization of the accreted stellar halo, which results in uncertainties and difficulties in inferring the merger history of galaxies. The extent of this population however varies dramatically from model to model, being dominant only in central regions < 11 kpc for some or up to 30 kpc for others \citep{pillepich2015,zolotov2009,cooper2015,wright2023} . These differences  motivate a detailed observational characterization of stellar halos that allows us to constrain the models and thus improve our interpretation of the accreted stellar halo and the merger history of a galaxy.

Unfortunately, reliable observations of stellar halos are extremely difficult to obtain due to their low surface brightness \citep[$\mu_V \ge 28\,\mathrm{mag/arcsec^2}$,][]{bullock2005}. Thus quantitative measurements of this component are highly challenging. 
For this reason, there are multiple approaches employed to measure and characterize stellar halos. Integrated light through deep imaging surveys can reach $\mu_V \sim 28-32\,\mathrm{mag/arcsec^2}$ and allows us to cover large areas of the sky while being observationally cost-effective \citep{martinez-delgado2010,vandokkum2014,merritt2016, Trujillo2016, Gilhuly2022}. However,  Galactic cirrus and scattered light are challenges that make this approach more suitable for galaxies at high latitude as well as for streams. 

On the other hand, observations using resolved stellar populations are more expensive, but allow us to reach fainter surface brightness \citep[$\mu_V \sim 32-34\,\mathrm{mag/arcsec^2}$][]{harmsen2017} than integrated light and avoids systematics like Galactic cirrus and scattered light.
In addition, resolved stars provide us an insight into the metallicity and age of the populations that make up the stellar halo. While ground-based wide-field telescopes allow one to build up a panoramic view of the stellar halos of galaxies \citep{ibata2014,okamoto2015,Crnojevic2016,smercina2020}, HST-based surveys can be done using pencil-beam fields directed at the stellar halos of galaxies out to 10 Mpc with exquisite star-galaxy separation \citep[e.g.,][]{rejkuba2009, Monachesi2013, peacock2015, cohen2020, Rejkuba2022}.  An example of the latter approach is that of the Galaxy halos, Outer discs, Substructure, Thick discs, and Star clusters (GHOSTS) survey, one of the largest HST programs designed to study resolved stellar populations in the outskirts of nearby edge-on disk galaxies to date \citep[][for a detailed description]{R-S11, monachesi2016a}. Targeting the red giant branch (RGB) stars along the minor axis as tracers of the accreted stellar halo population in the F606W and F814W filters using both the Advanced Camera for Surveys (ACS) and Wide Field Camera 3 (WFC3), the GHOSTS survey has revealed a striking degree of diversity in stellar halo masses, density profiles and color (as proxy of metallicity) for six MW-like edge-on galaxies \citep{monachesi2016a,harmsen2017}. The GHOSTS survey also derived a stellar halo mass-metallicity relation \citep{harmsen2017}, suggesting that the inner $\sim$50kpc of stellar halos for MW-like galaxies are dominated by a single massive accretion event \citep{deason2016, dsouza2018MNRAS, monachesi2019}. 

In addition to photometric studies, kinematic approaches provide an alternative and crucial tool for differentiating the stellar components of a galaxy, which is hard to achieve using solely photometric data. Kinematic signatures can help infer and unveil the merger history of galaxies as well as differentiate between the accreted and in-situ component of halos \citep{gilbert2007,gilbert2012,escala2019,gilbert2022,dey2023}. Naively, one would expect to observe no net rotation in the stellar halo of a galaxy as accreted satellites are expected to arrive randomly \citep{bullock2005}. However, the accretion of a single massive satellite could produce a rotational signature in the halo, which might even influence the alignment of the disk \citep{Gomez2017, dsouza&bell2018}. A rotational signature could also be produced by the contribution of disk stars that were kicked out and have now become part of the halo \citep{pillepich2015, monachesi2016b}. The properties of stellar halos along their major axes, in contrast to the minor axis properties, are particularly insightful in distinguishing between an accreted stellar halo or an in-situ dominated component:
while a rotation in the stellar halo along the major axis could be caused by a single large accreted event or kicked-up disk stars, the detection of zero rotation in the halo unambiguously points to it being of accreted origin.

Yet, measuring the velocity of stellar halos of MW-like galaxies outside the Local Group is extremely hard, due to the low surface brightness, which prohibits the analysis of standard integrated spectroscopy. Kinematics of halos in external galaxies have been performed using discrete tracers, such as planetary nebulae or globular clusters \citep[e.g.,][]{Pulsoni2020, Hartke2022, Pulsoni2023, Bhattacharya2023}, taking clear advantage of their brightness, but these are sparse sampling and are preferentially tracing special stellar populations compared to the bulk of the stellar halo population, which is indeed traced by the RGBs. Consequently, so far only the diffuse stellar halos of the MW and M31, i.e. only two MW-like galaxies, have been characterized kinematically, from spectroscopy of their resolved stars. While the MW has little-to-no rotation   \citep[$V_{rot}\sim 5-25$ \km,]{deason2017, helmi2018,bird2021,yang2022}, M31 has strong signatures of halo rotation, with V$_{rot} \sim 150$ \km out to beyond 40 kpc along the major axis \citep{ibata2005,dey2023}. The MW is thought to have accreted an LMC-like progenitor on a extremely radial orbit 8-11 Gyr ago \citep{Belokurov2018,helmi2018}, while M31 is thought to have accreted a large progenitor ($\log\, M_{*}\,\sim$ 10.3) 2-3 Gyr ago \citep{dsouza&bell2018,hammer2018}. Kinematic constraints of the stellar halos of MW-like galaxies outside the Local Group, and in particular of a typical stellar halo, in between the extreme stellar halos of MW and M31, can give us valuable insights into their formation mechanisms, as well as help us constrain the in-situ component of the stellar halo.

In this work, we present such a measurement: the velocity and velocity dispersion of the stellar halo of a MW-like galaxy NGC 4945 (located at a distance of 3.56 Mpc \citealt{monachesi2016a}) from its resolved stars. \camila{NGC 4945 is a typical MW-like galaxy with a stellar mass of 3.8$\times 10^{10}\solarmass$ \citep{harmsen2017} and a luminosity of $M_B = -20.5$ \citep[from HyperLEDA\footnote{\url{http://leda.univ-lyon1.fr/}};][]{makarov2014}. 
Both NGC 4945 and the MW are spiral galaxies, but NGC 4945 has a maximum rotation velocity of $\sim$180 \km, which is lower than than of the MW. While these galaxies are not identical in their properties, they can be considered similar within the realm of giant MW-like galaxies.} This represents the first resolved stellar kinematic measurement of a diffuse stellar halo beyond the Local Group. 

To measure the halo kinematics of NGC~4945, we present a technique that combines new deep Multi Unit Spectroscopic Explorer (MUSE) observations of a stellar halo field along with existing HST images. This combination enables us to utilize the accurate astrometric positions of our targeted stars from the HST catalogs to extract their spectrum from the MUSE datacubes. The technique we employ here is similar to that presented by \citet{toloba2016}, but has key differences (see Section~\ref{sec:methodology}).   \citet{toloba2016} spectroscopically targeted blends of RGB/AGB stars using individual slits (Keck Deimos spectrograph coupled with Subaru/Suprime-Cam imaging) in order to constrain the dynamics of NGC 4449  (at a distance of $\sim$4 Mpc) and its stellar stream. By further co-adding the spectra of the RGB/AGB blends, they were able to maximize the signal-to-noise ratio (S/N) of the final stacked spectrum to measure a radial velocity. 

\subsection{Why NGC~4945?}

NGC 4945 features a typical MW-mass stellar halo, with a relatively large stellar halo mass \citep[$\sim 3.5 \times 10^9 \solarmass$,][]{harmsen2017}, and an oblate (c/a = 0.51, \citealt{harmsen2017} at 25 kpc) and metal-rich ([Fe/H]$\sim -0.9$ dex, \citealt{monachesi2016a}) stellar halo, characterized using HST photometry of RGB stars as tracers for its underlying population. 
 \citet{monachesi2016a} also shows that NGC 4945 has a flat halo color radial profile, both on the major and minor axes, reflecting a flat metallicity profile.
These characteristics leave us at a crossroads. On one hand, we have a stellar halo that is large and oblate, suggesting a strong contribution from kicked-up disk star, shaping its flattened distribution. On the other hand, the similarity in color profiles along the major and minor axis hints at an accretion origin. Thus, only a kinematical measurement has the potential to disentangle its origin.
If we detect signs of rotation in NGC~4945 halo, this could be attributed to either a in-situ and accreted component. However, the absence of rotation would unequivocally indicate that its stellar halo is dominated by accretion. 

NGC 4945 is an optimal galaxy for our investigation. Positioned at a distance of 3.56 Mpc \citep{monachesi2016a}, this nearby edge-on galaxy similar to the MW not only permits the resolution of its stars using HST but also it can be observed with MUSE, due to its sky location. NGC 4945 hosts both a Seyfert 2-type AGN and a starburst. It is included in the Bright Galaxy Catalog of the HI Parkes All Sky Survey (HIPASS) \citep{koribalski2004} which determined a systemic velocity for this galaxy of 563 \km from its HI spectrum. A newer study of HI gas in this galaxy was presented by \citet{ianjama2022}. They studied the neutral gas using MeerKAT data, which are more sensitive than the previous HI work by \citet{koribalski2018}, showing that its HI disk is slightly larger than the bright optical disk. They also found a large amount of halo gas around the disk with an asymmetric distribution, with more HI located through the receding side of the galaxy. This halo gas does not follow the kinematics of the rotating disk and has velocities closer to the systemic velocity. The authors suggest that this gas is likely due to outflows driven by the central starburst of the galaxy \citep[as seen in NGC 253, ][]{lucero2015}{}{}. 

In this work, we target two fields along the major axis of NGC~4945: a halo field at 34.6 kpc and an outer disk field at 12.2 kpc which serves as a control field. We already have existing HST observations from the GHOSTS survey, and  we aim to measure the mean velocity and velocity dispersion in each of these fields by co-adding the spectra of the individual RGB and AGB stars. This work presents the first kinematical measurements of a diffuse stellar halo from its individual stars in galaxy outside the Local Group.

The paper is organized as follows. Section 2 describes the HST photometric and MUSE spectroscopic data. Section 3 explains the methodology developed to obtain the final spectra and how the velocity measurements were obtained. Section 4 presents the main results of this work. In Section 5 we discuss our results and place them into context with the results from the MW and M31 stellar halos as well as with cosmological simulations. Finally, in Section 6 we summarize our conclusions.

\section{Observations} \label{sec:obs}
Our goal is to measure the radial velocity and velocity dispersion of two fields along NGC 4945 major axis, using resolved stars. To accomplish this, we analyze two newly acquired MUSE datacubes, in combination with existing GHOSTS photometric catalogs from HST observations, which provide precise positions for the individual NGC 4945 stars.  The MUSE fields are strategically positioned at distances of 12.2 kpc and 34.6 kpc from the galaxy's center along its major axis. These locations were chosen to specifically measure the galaxy's outer disk and halo populations, respectively. Subsequent subsections provide detailed descriptions of both the photometric and spectroscopic datasets.

\subsection{Photometric Catalog from GHOSTS: selection for MUSE fields}\label{sec:ghosts}

The GHOSTS survey\footnote{\url{http://vo.aip.de/ghosts/index.html}} provides catalogs of stars in each imaged fields. Each galaxy in the survey has several HST pointings observed with the Advanced Camera for Surveys (ACS) and the Wide Field Camera 3 (WFC3) along the major and minor axis. The photometric catalogs of each of these pointings were obtained using the GHOSTS pipeline, described in detail in \citet{R-S11} and \citet{monachesi2016a}. These provide the positions of the detected sources, their magnitudes calibrated on to the VEGAmag HST system and parameters used to discriminate between stars and other sources like background galaxies or cosmic rays. In this work we use the final clean catalog of stars, i.e. already decontaminated for unresolved background galaxies, which was presented in \citet{monachesi2016a}. We refer the reader to that work for details regarding the construction of the star catalogs for NGC~4945 fields.

\begin{table}
	\centering
	\caption{Positions (RA and Dec) of both MUSE FoV, the total integration time of the final datacubes, FWHM and the distance of the field to the center of the galaxy.}
	\label{tab:fields}
     \setlength{\tabcolsep}{3pt}
    \begin{tabular}{lccccc} 
		\hline\hline
            \noalign{\smallskip}
		Field & RA & Dec & $t_{\rm exp}$ & FWHM & D \\
		&(h:m:s)  &(d:m:s)  & (s)& (arcsec) & (kpc)\\
		\hline
             \noalign{\smallskip}
		Halo  & 13:07:41.64 & -49:02:54.74 & 29920 & 0.7 & \camila{34.6}\\
		Outer disk & 13:06:17.66 & -49:19:40.25 & 2520 & 0.7 & \camila{12.2}\\
		\hline
	\end{tabular}
\end{table}

\begin{figure*}
\centering
\begin{subfigure}[l]{0.6\linewidth}
  \includegraphics[width=\linewidth]{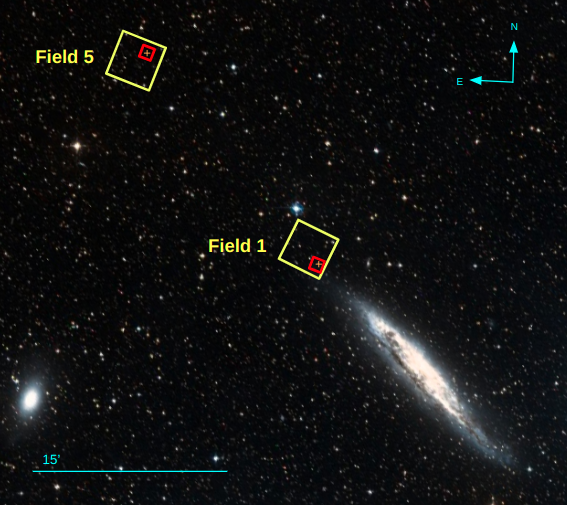}
\end{subfigure}
\begin{subfigure}[c]{0.3\linewidth}
  \begin{tabular}{c}
    \includegraphics[width=2.0in,height=2.0in]{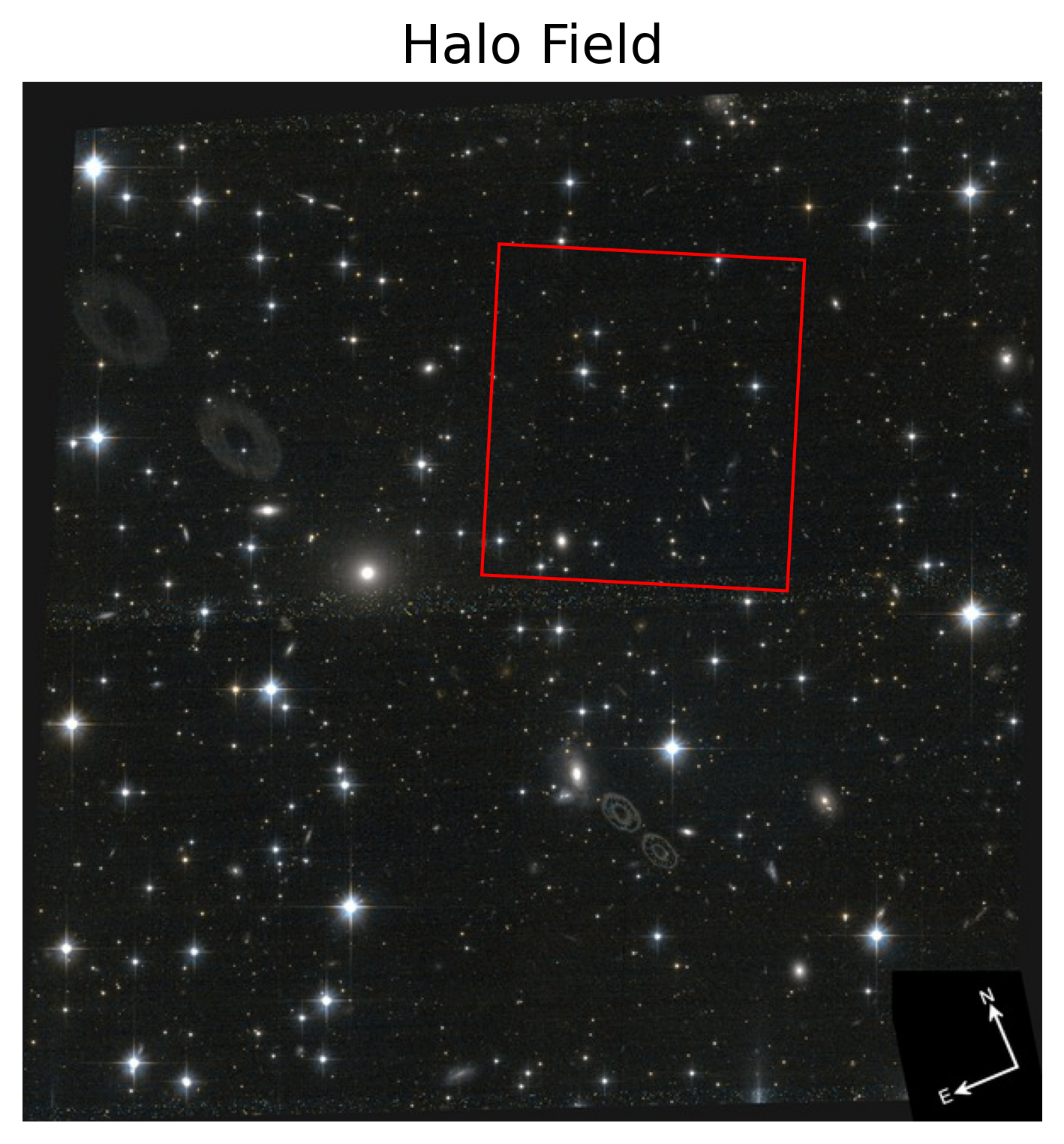} \\
    \includegraphics[width=2.0in,height=2.0in]{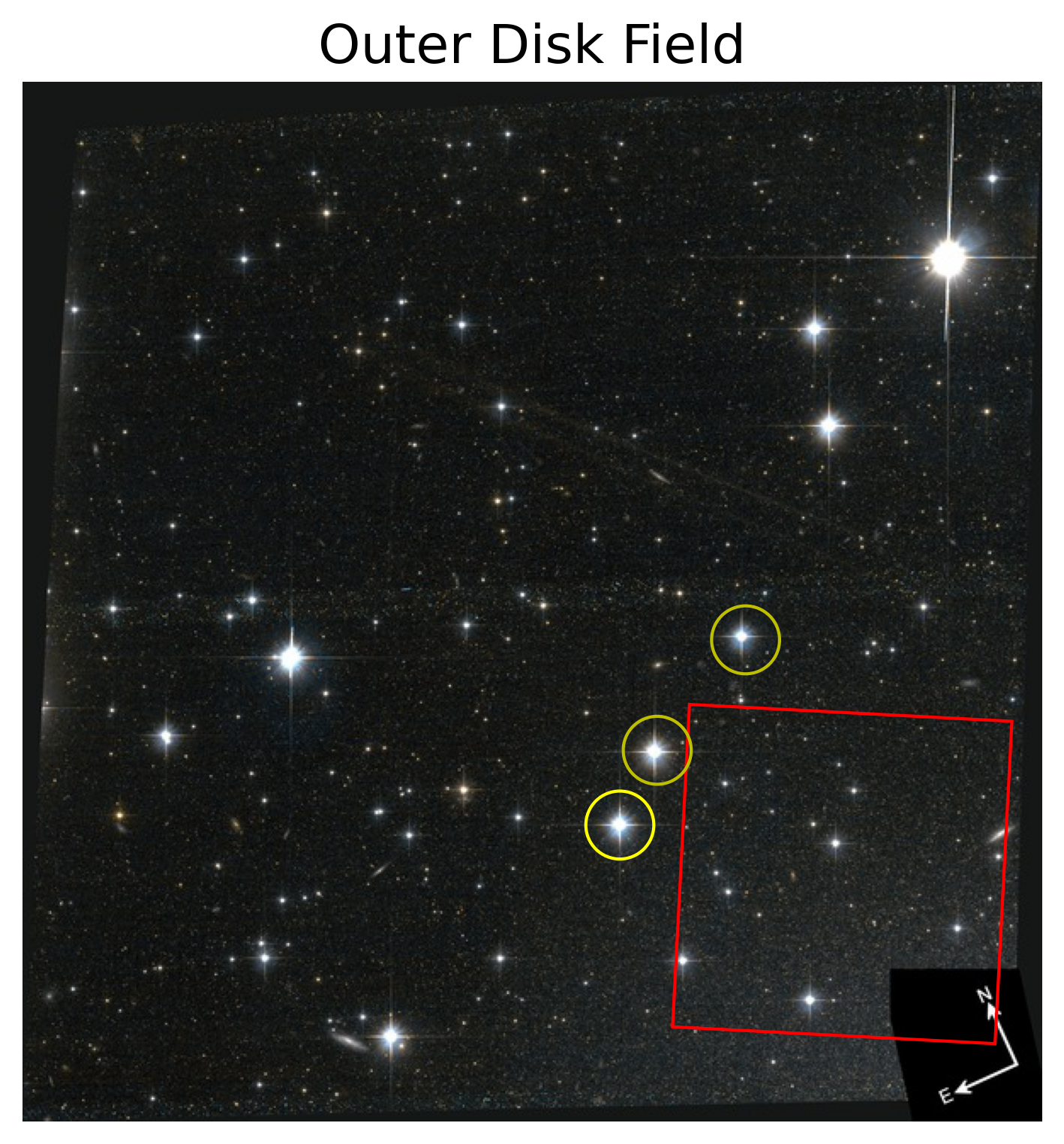} 
  \end{tabular}
\end{subfigure}
\caption{(Left) DSS colored image of NGC 4945 showing the location of the GHOSTS Field 1 and 5 (yellow boxes) and the MUSE targets (red boxes). Field 5 is at $\sim$35 kpc and Field 1 is at $\sim$15 kpc from the center of the galaxy along the major axis. (Right) HST ACS/WFC images of Field 1 (bottom panel) and 5 (top panel). The FoV of each image is  3.4'$\times$ 3.4'. Red boxes show both MUSE FoVs (1'$\times$1') inside the HST ACS/WFC Fields 1 and 5, and yellow circles highlight the 3 bright MW stars that are outside the Outer disk FoV but close to one edge of the FoV.}
\label{fig:Fields}
\end{figure*}

Of the several GHOSTS fields observed in NGC 4945, we focus in this work on two of them, Field 1 at $\sim$15 kpc (outer disk) and Field 5 at $\sim$35 kpc (halo) observed with the ACS along the major axis of this galaxy. 
The choice of the major axis was motivated by the fact that models predict that the contribution of kicked-up disk stars, i.e. the in-situ halo, is more noticeable along the major axis \citep{monachesi2016b, Gomez2017}.
Additionally, as the size of the MUSE field is smaller than the ACS/HST field, special care must be taken to maximize the number of RGB stars with spectroscopic counterparts. We choose our MUSE pointings within the two fields - such that we maximize the number of RGB stars while at the same time avoiding bright MW stars. The distance of the halo field was chosen so as to reach as far out as possible into the stellar halo but at the same time to have enough RGB spectra to stack be able to reach the required S/N (see Sec.~\ref{sec:muse}). The GHOSTS field at 40 kpc is the farthest field along the major axis showing a well populated RGB. The MUSE field for the outer disk is strategically positioned in one corner of the GHOSTS FoV. This placement is intentional, aiming to maximize the contribution from the disk. As mentioned earlier, this region will serve as the control field for velocity comparisons.

The left panel of Fig.~\ref{fig:Fields} shows the positions of the 2 GHOSTS fields in yellow and the two MUSE pointings, within those fields, in red. 

\subsection{MUSE Spectroscopic Data}\label{sec:muse}

\begin{figure}
    \centering
    \includegraphics[width=\columnwidth]{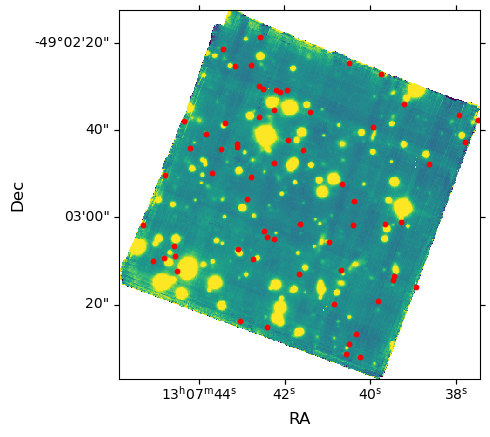}
    \includegraphics[width=\columnwidth]{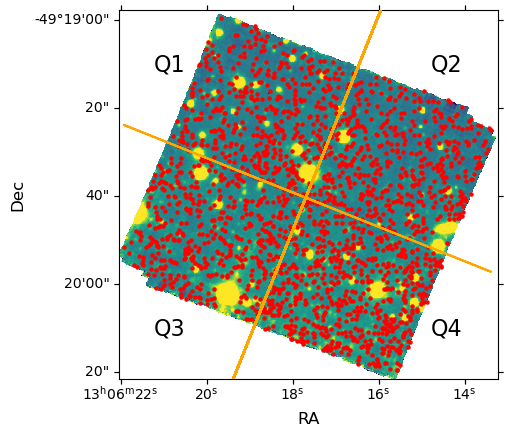}
    \caption{Halo (top) and outer disk (bottom) MUSE FoV. The images were created using the flux in the wavelength range from 6869 to 9300 \angstrom, representative of the F814W HST/ACS filter. Red points are the RGB and AGB+RGB stars, until two magnitudes below the TRGB selected by GHOSTS in each FoV.}
    \label{fig:FoV}
\end{figure}

Observations were taken in January, February and March, 2020, and in July, 2019, in service mode during dark time and photometric conditions with an average seeing of $0.8''$ (observing program 103.B-0514, PI: A. Monachesi). The instrument was set up in wide-field mode, without adaptive optics. 
We use the instrument in its nominal-wavelength mode (480-930 nm). This setup ensures a FoV of $1' \times 1'$, a spatial sampling of $0.2''$ per pixel, with a spectral sampling of 1.25 \angstrom/pixel. Each resolution element is sampled by 2.5 pixels along the spectral direction, which corresponds to a velocity resolution of 50-80 \km. The spectral resolution of MUSE at 9300 \angstrom is 3590. The choice of the wide-field mode allows us to cover a larger area and thus include as many stars as possible.

The integration time was chosen such that it would be possible to reach the required minimum S/N>2 per resolution element in the final stack to measure reliable velocities. The two GHOSTS fields 1 (outer disk) and 5 (halo) have a surface brightness of $\sim$26 and $\sim$29 mag/arcsec$^2$, respectively \citep{harmsen2017}. We expect to find 2000 and 150 RGB stars for co-addition in the respective MUSE FoV. Taking into consideration these numbers,  the halo field was observed for 8 h in 11 Observing Blocks (OB) of 2720 s each. Each OB consists of 2 exposures of $\sim1320$ s and a rotation of $90^\circ$. The outer disk field was observed for 42 min in 1 OB of 4 exposures of 630 s. Table~\ref{tab:fields} lists the positions of both fields, the integration times, FWHM and distances from NGC 4945 galactic center. 

All the data have been reduced by the MUSE consortium using the official pipeline version 2.8 \citep{weilbacher2020}. All the exposures were processed by the  MUSE scibasic recipe which used the corresponding daily calibrations (flatfields, bias, arc lamps, twilight exposures) to produce a pixel table containing all pixel information: location, wavelength, photon count and an estimate of the variance. The pipeline recipe scipost is then used to perform astrometric, flux calibrations and the sky subtraction on the pixtable to create the final reduced data cubes. We use the two final integrated data cubes, one per field, containing a flux- and wavelength-calibrated spectrum in each spaxel as a starting point\footnote{The data are publicly available \url{http://archive.eso.org/cms.html}}, to which we then make an astrometric correction and additional emission sky lines subtraction, as we show in Section~\ref{sec:methodology}, before the spectral analysis. 

In Figure \ref{fig:FoV}, we show the images of the MUSE datacubes extracted within the HST F814W passband, i.e. using the flux in the wavelength range from 6869 to 9300\angstrom, for the outer disk and halo MUSE fields.

\subsection{Color-magnitude diagrams of stars in the MUSE fields}\label{sec:muse_cmd}

\begin{figure*}
    \includegraphics[width=2\columnwidth]{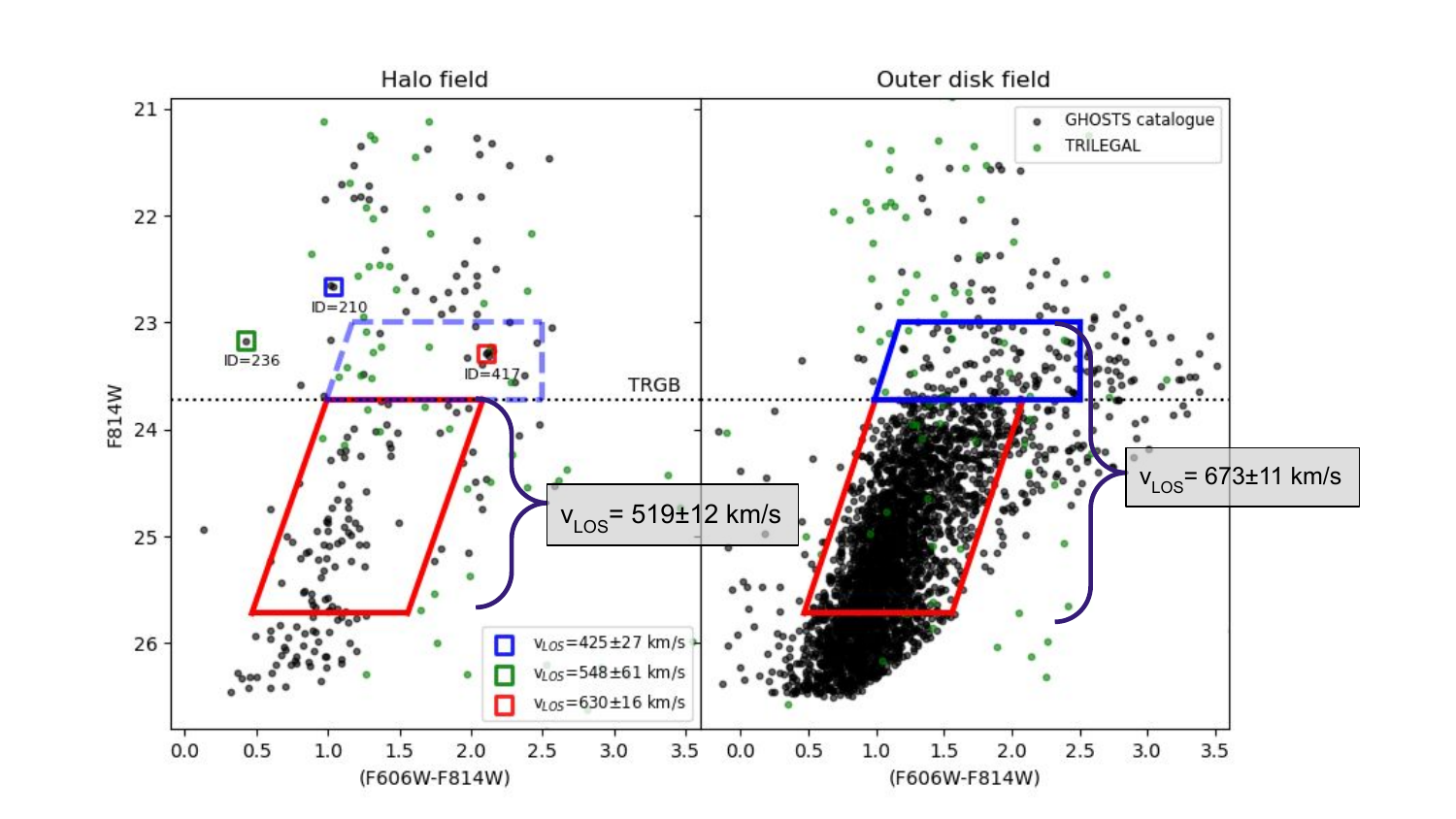}
    \caption{CMDs of detected stars from GHOSTS in the MUSE Halo (left) and Outer Disk (right) FoV. Magnitudes are corrected for Galactic extinction. The magnitude of the TRGB is 23.72 in the F814W filter and it is marked with a black dotted line. Green points represent the contamination from MW foreground stars estimated using the TRILEGAL model. The red and blue boxes are illustrative and indicate the regions in the CMD where RGB and AGB stars are selected \camila{respectively}, both in the halo and disk fields. In the halo field, for stars brighter than the TRGB, the S/N allowed for individual radial velocity measurements. Three of these stars, highlighted \camila{with the small green, blue and red boxes}, have radial velocities greater than 400 \km, which is the minimum velocity threshold we set to identify stars belonging to NGC 4945. Additionally, one of them falls within the AGB box (dashed light blue line). We marked this box differently from the one in the disk field to indicate where AGB stars would mainly be located in this field. However, these AGB stars are not added to the final stack spectrum; only the RGB stars are included. In the case of the outer disk, since the contamination from MW background stars is negligible compared to the number of NGC 4945 stars, both RGB and AGB stars were used in the final stack.}
    \label{fig:cmds}
\end{figure*}

In Fig.~\ref{fig:cmds} we show the GHOSTS color-magnitude diagrams (CMDs) of the MUSE regions: the halo (left) and outer disk (right) containing a total of 184 and 2975 stars respectively. The magnitude of the tip of the RGB (TRGB) for NGC 4945 was calculated by \citet{monachesi2016a} to be 23.72 in the F814W filter. These CMDs reach two magnitudes below the TRGB, well above 50 completeness level (see \citealt{monachesi2016a}), and they clearly evidence the presence of a well-defined RGB population (both in the halo and outer disk field) and AGB stars (in the outer disk field). The RGB stars are the main sources for characterizing the halo field, while AGB stars are additionally incorporated to characterize the outer disk field. Also in Fig.~\ref{fig:cmds}, we illustrate the delimited region for selecting RGB stars (red box) and AGB stars (blue box), with F814W magnitudes between 23 and 23.72 and a color range between 0.8 and 2.5. Figure \ref{fig:FoV} displays, as red points, the distribution of the selected RGB stars within the red box for the halo MUSE field and the distribution of the selected RGB and AGB stars (blue and red box) for the outer disk MUSE field.
The boxes that we use for analysis in this work have been predominantly defined using the CMD of the outer disk field. We adopted a conservative red limit to the AGB box to avoid as much as possible contamination from MW stars (see below). The astrometric precision of the GHOSTS data aids in the extraction of the spectra of individual RGB and AGB stars in each field. 

In addition to the dominant population of RGB stars in the halo of NGC 4945, owing to its low Galactic latitude $b=13.3^{\circ}$, the CMDs should also contain MW foreground stars as contaminants. We estimated the contamination level of foreground stars using \textsc{trilegal} models\footnote{\url{http://stev.oapd.inaf.it/cgi-bin/trilegal}} and it is shown in Figure~\ref{fig:cmds} as green dots. Most of the MW contaminants are stars brighter than the TRGB or redder than the RGB ($F606W-F814W>1.6$) and has a higher fractional contribution in the low-density halo field. If few of the stars above the TRGB are truly NGC 4945 stars in the halo field, it is important to find them, since those would be the brightest stars in our sample, thus they will contribute significantly to the combined spectrum in that field. Thanks to the large integration time in the MUSE halo field, individual spectra for stars brighter than the TRGB (i.e. with $F814W<23.72$) have enough S/N to measure their velocities, which permits a discrimination between NGC 4945 ($400 < v_{los} <800$ \km) and MW ($v_{los}<400$ \km) stars. In the outer disk field, it is not possible to obtain reliable velocity measurements of individual NGC 4945 stars, due to the shorter exposure time. Nevertheless, given the higher stellar density (see Fig.~\ref{fig:FoV}) the number of MW contaminants is negligible compared to the AGB and RGB stars in that field, thus there is no need to kinematically discriminate the MW contaminants from the stack spectrum of the outer disk field.

\section{Methodology} \label{sec:methodology}

In this section, we describe the procedure used to extract individual star spectra from our MUSE observations, the stacking process and the procedure for measuring the line-of-sight (LOS) velocities and velocity dispersions. 

Since our target stars are fainter than 23 in I-band magnitude, their individual S/N range between 0.5-4 and 0.1-1 for the halo and the disk RGB respectively. For this reason, we were unable to utilize the standard stellar spectra extraction software PampelMUSE \citep{Kamann2013}. PampelMUSE was developed to extract stellar spectra in crowded fields optimizing for the PSF as well as the fluxes of de-blended stars and works very well in the high S/N regime (S/N >10). On the other hand, in the low S/N regime, it becomes very difficult to optimize for the flux of a star.  Instead, in the limit of working at very low S/N, we develop our own custom spectral extraction method akin to "forced PSF photometry", which allows us greater flexibility.  Instead of optimizing for the Point Spread Function (PSF), we constrain it from the very bright stars present in both our MUSE FoVs. Given a PSF, we then extract the spectra of our target star at their proper locations forcing the target to have the PSF. 

For the kinematical measurements, our approach is based on the one developed by \citet{toloba2016} but has key differences. We use higher-resolution HST imaging (0.05’’ per pixel), allowing us to target fainter individual stars and to perform a more precise astrometry. Additionally, MUSE's IFU capability lets us observe all stars in its field of view and use surrounding pixels as local background. The position of Toloba's slits are mainly within the stream of NGC 4449, which has higher stellar density and surface brightness \citep[$\mu_g=26.72\,\mathrm{mag/arcsec^2}$,][]{martinez-delgado2012} compared to the surface brightness of the stellar halo at 40 kpc from the center of NGC 4945 \citep[$\mu_V\sim 29\,\mathrm{mag/arcsec^2}$,][]{harmsen2017}{}{}, where one of our MUSE fields is located. 
While MUSE has lower spectral resolution (R$\sim$3000) than Toloba et al.'s data, its broader spectral coverage allows to fit a wider wavelength range and cover more spectral features.
Therefore, MUSE in combination with HST imaging, offers a new and powerful way of measuring the velocity field of the stellar halos of nearby MW-mass galaxies outside the Local Group.

In the following subsections, we describe our astrometric corrections to the MUSE datacubes, how we constrain the PSF, our stellar extraction procedure and finally, the stacking process and procedure to measure the velocities, velocity dispersions and their associated uncertainties.

\subsection{Astrometry and Sky subtraction} \label{sec:astrometry_zap}
To accurately extract the stellar spectra of our target stars, we must have a good astrometric correspondence between the MUSE fields and the corresponding HST images. Thus we align the WCS of both MUSE fields with the corresponding HST images from GHOSTS. This step is crucial, as the MUSE pipeline's default astrometric alignment of the data exhibits a shift with respect to the  calibration provided by HST. To perform the astrometric correction for each datacube, we employed the \textsc{mpdaf}\footnote{\url{https://mpdaf.readthedocs.io/en/latest/index.html}} tool.

The standard sky subtraction procedure carried out by the MUSE pipeline leaves behind residual sky lines within the final data cube. To improve the sky subtraction in our datacubes, we employed the post-processing tool Zurich Atmosphere Purge \citep[\textsc{ZAP};][]{zap}, which effectively removes any residual sky emission lines.
The ZAP code uses a principal component analysis to enhance sky subtraction while preserving the shape and flux of astronomical source lines, and relies on obtaining a pure sky spectrum as an input. We obtain this by masking out all the sources ($1\sigma$ above the global background) using the \textit{photutils} tools. 
We further manually mask out all the sources in the GHOSTS catalog with magnitudes down to F814W$\sim$26.5, which were not detected by \textit{photutils}. 
For the halo field, we use a masking radius of 7 spaxels for every GHOSTS source, given the lower star density in this region. Conversely, for the outer disk field, where the star density is higher, a smaller masking radius of 1 spaxel was employed. 
The masking procedure leaves us with 42894 and 46250 spaxels for the halo and outer disk field, respectively, which represent a 40.1 and 42.9 percent of the available spaxels in each datacube. With the constructed sky mask, we applied the ZAP algorithm with its default values to construct a sky residual spectrum for each field. 
This spectrum was then subtracted from the original data cube, effectively eliminating the majority of residual sky lines that contaminate the dataset. 

We also assess the impact of a more restrictive masking by enlarging the radius around each masked source to 10 spaxels, leaving then fewer spaxels available for calculating the sky spectrum. We compare the standard deviation of the resulting clean sky spectra using both masks and find that applying a more restrictive masking did not yield a noticeable improvement in removing the residual sky lines.  

\subsection{Point Spread Function} \label{sec:fsf}

An accurate description of the PSF is essential for an optimal extraction of a stellar spectra from a datacube, especially when dealing with faint low-magnitude stars. We use the MUSE specific tools \textsc{mpdaf} to create a wavelength-dependent PSF model, which we assume to be spatially constant throughout the entire FoV. This model employs a Moffat circular function to describe the radial shape of the PSF, with parameters such as FWHM and $\beta$ varying with wavelength. To build this model, we select an isolated bright MW star within a 90x90 spaxels sub-cube centered around it. We then mask out all nearby sources. We fit the PSF model on 20 evenly spaced layers along the wavelength axis, adjusting the FWHM and $\beta$ parameters accordingly. This process results in a PSF cube that encompasses the entire wavelength range. In addition of using the PSF to extract the spectra of our target stars, it is also utilized to create a model of the bright MW stars in our FoVs to estimate the contamination level from them on our target fainter stars, as we show in the next subsection \ref{sec:contamination}.

\subsection{Contamination from bright and neighboring sources}\label{sec:contamination}

In our final analysis, it is important to ensure that the individual stellar spectra contain as little contamination as possible, specially in the halo field where the number of RGB stars is significantly lower compared to the outer disk field. 
Various sources of contamination may affect the light of the spectra of interest. Firstly, scattered light from bright MW stars, particularly those with magnitudes above the TRGB, contaminates the FoV and makes it difficult to extract clean spectra from fainter stars. 
Additionally, fainter stars comparable in brightness to our target RGB stars, which are redder than the RGB box defined in Fig.~\ref{fig:cmds} (and hence likely MW star), may also contaminate the extracted flux of our sample of RGB stars. Finally, extended galaxies that are near in projection to our RGB stars may also introduce contamination to their individual stellar spectra. 

To assess which stars in our RGB and AGB selected boxes (see Figure~\ref{fig:cmds}) from the GHOSTS catalog are affected by these potential sources of contamination, we create a 2D model of the contaminants in both fields.
In each of the MUSE FoVs, there are several bright stars (virtually all the sources that appear as yellow in Figure~\ref{fig:FoV}) that are not present in the cleaned GHOSTS catalog since they are saturated or fragmented into small entities by the GHOSTS pipeline. Some of these stars have been identified by Gaia Survey \citep{gaiaedr3}, totalling 18 and 15 in the Halo and Outer disk, respectively.
We utilize \textit{photutils} to identify stars and galaxies within each MUSE white-image field. In the halo field, we identify a total of 125 bright stars, 7 foreground galaxies, and 19 faint stars outside the RGB box. In the outer disk, we identify 52 bright stars and one foreground galaxy. Subsequently, we construct a mock halo/outer disk image of the FoV, respectively, by positioning a scaled and wavelength-flattened PSF for each individual contaminating stars identified.
Furthermore, we modelled the projected galaxies with a two-dimentional Sersic function whose parameters (Sersic index, effective radius, position angle, and central coordinates) we extracted from the image using \textit{galfit} \citep{galfit2002}. 

We estimate the projected flux of scattered light from these sources at the position of each star from the GHOSTS catalog within the RGB and AGB boxes selected for our analysis. We then compare this estimated flux with the actual projected flux of these stars at their central spaxel and calculate the percentage of contamination that reaches each star. Given that we are working with very faint stars, we conducted several tests to evaluate different levels of contamination (5 to 30\%) to minimize the inclusion of light from neighboring sources in our stacked spectrum of RGB/AGB stars. 
We find that the velocity measurements are stable in all cases. Thus, to maximize the number of stars in each field, and thus the S/N in the stacked spectrum, and at the same time minimize the contamination as much as possible, we will include in our analysis stars with contamination levels up to 20\% for the halo field and up to 30\% for the outer disk field. This corresponds to 53 and 1021 RGB stars in the halo and outer disk field, respectively, when considering stars down to two magnitudes below the TRGB.

\subsection{Star Spectrum Extraction procedure and background subtraction}\label{sec:extraction}

In this section, we explain our approach to extracting the spectra of RGB and AGB stars in both the halo and outer disk fields, focusing on maximizing their S/N while minimizing contamination from the background sky. 
Because of the differences in depths and crowding of the fields, different procedures were employed in the halo and outer disk field for estimating the sky background and for the extraction of the star spectrum.  

The extraction was guided by the following steps:
In the halo field, we compute the local sky background surrounding each star by selecting a square region. The size of this region depends on the star's magnitude, ranging from a square of 31 spaxels per side for brighter stars to 13 spaxels per side for those stars fainter than the TRGB. 
Within each of these squares, all detected sources, including the target star itself, are masked. We emphasize that this masking process aims to exclude any potential contamination, ensuring that only uncontaminated spaxels contribute to the construction of the local background spectrum for each star. The local background spectra are then created by taking the median of the available spaxels (i.e. unmasked) in each selected square region. 
After determining the locally calculated background spectrum for each star, we move on to the next step: extracting the spectrum of each star.  We extract the mean PSF-weighted spectrum for each star with different extraction radius based on the star's magnitude, ranging from 4-spaxel radius (52 spaxels in total) for the brightest stars to a 1-spaxel radius (4 spaxels in total) for stars fainter than the TRGB. The PSF-weighted spectrum extraction, coupled with the adaptive radius, ensures precise extraction of spectral information from stars across different brightness levels. We then subtract the locally calculated background spectrum to each of the extracted target star's spectrum. It is worth noting that there is a gap in radius of 2 pixels between the extracted region of the target star and the region used to calculate the local sky background around each target star. For stars fainter than the TRGB this gap is maintained at 2 pixels, while for brighter stars above the TRGB, the radius increases to 4-6 pixels.
This minimizes the possibility of oversubtraction, i.e. of including some flux from the target star itself into the local background estimate, reinforcing the purity of the extracted spectra.

For the outer disk field, we introduce two modifications to the extraction procedure due to the high density of stars and the limited availability of sky spaxels for calculating a reliable local sky background. Firstly, we extract the PSF-weighted spectrum using only 4 spaxels centered on each target star. Secondly, we adopt a global sky background. For this purpose, we divide the outer disk field into four equal sized quadrants as shown in Fig.~\ref{fig:FoV} (Q1 and Q2 are in the upper-left and right, while Q3 and Q4 are in the lower-left and right) and assume that the sky background remains constant within each quadrant. We then estimate the sky background in each quadrant by calculating the median spectrum of the source-free spaxels, following the mask created for ZAP. 
Then, we subtract the quadrant-determined sky background spectrum from each extracted star spectrum, based on its quadrant location.
However, the number of source-free spaxels varies significantly between quadrants. Specifically, Q4, located closest to the galaxy disk has half the number of source-free spaxels than Q2.  Consequently, due to the high crowding in Q4 and hence the rather unreliable determination of the sky background (4582 spaxels to estimate the background), we adopt the sky background estimated in Q2 for stars in Q4. 
This methodological adjustment, i.e. tailoring the background subtraction to the specific conditions of each quadrant, ensures a more reliable background subtraction in the outer disk field, enhancing the quality of the extracted spectra.

As an additional test, we also use the corresponding sky background spectrum calculated in Q4 for the target stars in Q4. The results obtained for the S/N and radial velocities for the stack spectra using this approach are quite similar (consistent within the errors) from those obtained by using the fiducial background subtraction. 
Nevertheless, given the higher crowding in Q4 due to its proximity to the disk, the likely presence of unresolved stars from NGC 4945 in that quadrant may contribute to the background there, which may result in a background spectra containing information from the disk population. Subtracting this information from each star's spectrum could lead to information loss. In turn, we adopt the more conservative approach, and decided to use the background from quadrant Q2 for the target spectra in quadrant Q4.

\subsubsection{S/N for individual star's spectra}\label{sec:snr}

Once we extract each star's spectrum, we measure its S/N.
Since the halo field is made up of 11 OBs, the final stacked cube has two problems that we have to deal with. First, there is a loss of S/N around the edges of the final stack depending on the number of OBs used to make up a certain spaxel. 
Therefore, in both fields we exclude stars close to the edge of the FoV: we enforce a criterion that requires stars to maintain a minimum distance of 6 spaxels from the edge of the FoV. 
Secondly, the variance of the final MUSE spectrum contains correlated noise introduced in the stacking process, due to the resampling of spaxels of the various observation blocks (OBs) \citep{bacon2107}.
We therefore decided not to use the variance as the noise estimate. Instead, we estimate the S/N directly from the star's extracted spectrum using the code DERSNR\footnote{ \url{http://www.stecf.org/software/ASTROsoft/DER_SNR/der_snr.py}} \citep{Stoehr2008}, in the wavelength range 8400-8800 \angstrom, which includes the CaT region.  
In short, this method utilizes the median value of the flux within the selected spectral region as the signal (S) and the third order of the median absolute difference (MAD) to estimate the noise (N), which is calculated as the median of the absolute differences between the flux at a specific pixel and the fluxes two pixels ahead and two pixels behind that given pixel. We assume in this process that the flux is a Gaussian function distributed and uncorrelated in wavelength bins spaced two pixels apart. We note that this approach is less accurate than when the full instrument and detector knowledge is taken into account, but can be uniformly applied across a single stellar spectrum as well as stacked spectra and has been widely used to do kinematics analysis of galaxies \citep[e.g.,][among others]{finlez2022,dago2023,comeron2023}.

\begin{figure}
    \centering
    \includegraphics[width=\columnwidth]{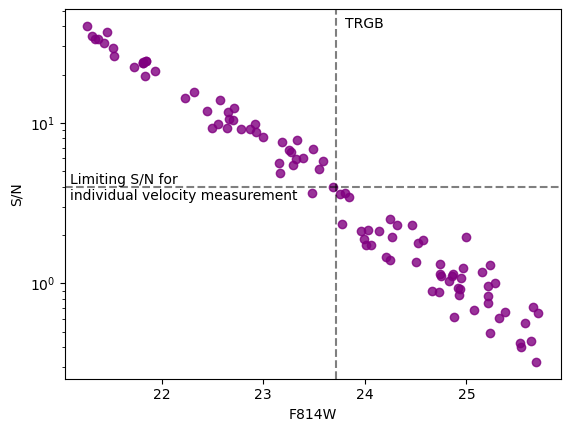}
    \includegraphics[width=\columnwidth]{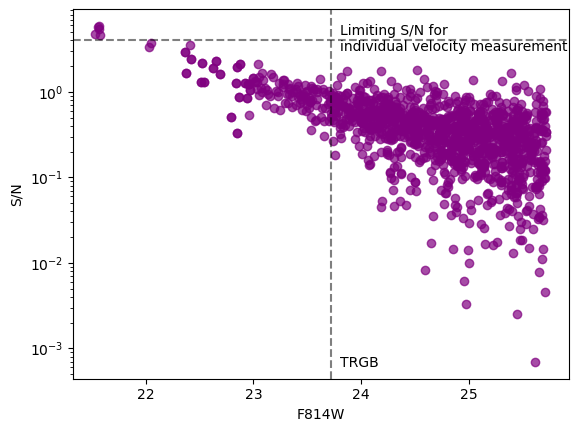}
    \caption{S/N per resolution element for each extracted spectra in the halo (top) and outer disk (bottom) fields measured in the CaT region 8400-8800\angstrom. The vertical line in the upper plot indicate the magnitude of the TRGB defined as the limit magnitude to measure individual velocities.
   }
    \label{fig:snr}
\end{figure}

In Fig.~\ref{fig:snr} we show the resulting S/N per MUSE resolution element (2.5 \angstrom) of each spectrum from uncontaminated stars (Sec.~\ref{sec:contamination}) selected in the red and blue boxes (see Fig..~\ref{fig:cmds}) for the halo (top panel) and outer disk field (bottom panel), as a function of their magnitude. The vertical lines mark the magnitude of the TRGB.  At this magnitude we obtained a S/N $\sim 4$ per resolution element for the halo field.  For magnitudes fainter than the TRGB in the halo field, as we demonstrate below in Sec.~\ref{sec:ppxf}, the S/N is not high enough to measure reliable individual radial velocities. For the outer disk field, the individual S/N obtained for each star is even lower. Consequently, the resulting spectra are very noisy, making it challenging to discern critical spectral features such as the CaT lines.

\subsection{RGB and AGB stars spectral co-addition} \label{sec:rgb_stacked}

To increase the S/N, we perform the co-addition of the individual star spectra, weighted by their corresponding S/N. This strategy assigns greater importance to the spectra with higher S/N, which contribute more signal.
Specifically, within the halo field, we co-add the spectra of the RGB stars in the selection box, encompassing the region from the TRGB down to two magnitudes below. In the outer disk field, in addition to the RGB stars, we also stack the 70 AGB stars within the selection box illustrated in Fig.~\ref{fig:cmds} to increase the final S/N of the stacked spectrum.
We generate four stack spectra for each field reaching down to different limiting magnitudes. Table~\ref{tab:combine_halo} outlines the magnitude ranges, the number of stars employed for each combined spectrum, and their corresponding S/N reached in the final halo field stack.  Similarly, Table~\ref{tab:combine_disk} lists the same details for the outer disk field.

\subsection{pPXF: measuring and testing velocities}\label{sec:ppxf}

Besides measuring the LOS velocities of the final stack, we also need to discriminate MW stars in the low density halo field from possible AGB stars belonging to NGC 4945 using their individual velocities.  To derive the LOS radial velocities, we use the penalized pixel-fitting method \citep[pPXF;][]{cappellari2004,cappellari2017} to perform a full spectrum fitting, which consists of modelling an observed spectrum by a best-fitting linear combination of differently weighted stellar templates. pPXF has been extensively used to extract stellar or gas kinematics and stellar population from absorption-line spectra of galaxies \citep[e.g. when S/N is 20,][]{boardman2017}. The pPXF method requires several inputs:  a set of template spectra, the noise spectrum and the starting value for the velocity and velocity dispersion $\sigma$ (the MUSE instrumental velocity dispersion is approximately 50 \km). We used the MUSE spectral library as stellar templates \citep{ivanov2019} to do the fit in a wide wavelength range. This set consists of 35 standard stars observed with MUSE covering stellar temperatures from 2600 and 33000 K, $\log g = 0.6-4.5$ and [Fe/H] from $-1.22$ to $0.55$. As a noise spectrum we use the variance of the pure sky spaxels of each field (the same as those used for ZAP). 
We vary the starting value for the velocity and its dispersion from 0 to 1000 \km, as we show below. 

\begin{figure}
    \centering
    \includegraphics[width=\columnwidth]{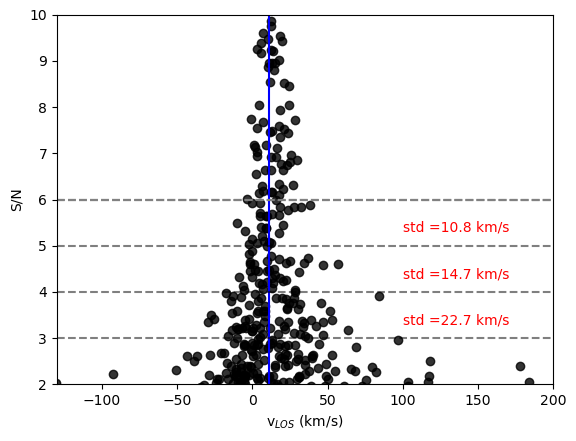}
    \caption{The scatter in the measurement of the velocity as a function of S/N for individual stars. We use an isolated bright star in the halo field, with F814W=22.58 and S/N=14 as template. The vertical blue line is the radial velocity of the star obtained with pPXF of 12.2 \km. In red we report the standard deviation at various levels of S/N, which decreases as the S/N increases.}
    \label{fig:noise}
\end{figure}

We need to evaluate the confidence of our results in order to be able to distinguish MW stars from possible stars belonging to NGC~4945 above the TRGB in the halo field. Given that our observations are in a very low surface brightness and S/N regime, we performed two tests to check the stability and reliability of the velocity measurements obtained with pPXF for our particular data set, when measuring individual star velocities.  First, we estimate the minimum S/N required to obtain reliable velocities using the entire wavelength range from 4800-8800 \angstrom. Second, we evaluate the dependence of the resulting velocity on the starting value for the velocity used as input in pPXF.

To measure the minimal S/N, we selected a relatively bright and isolated MW star within our FoV with a magnitude F814W=22.58, a S/N$\sim$14 per resolution element and a measured LOS velocity of this star is 13 \km obtained from a full spectrum fitting in the entire wavelength range. Then, we progressively add random Gaussian noise the the stellar spectrum: the magnitude of the perturbations is increased from a standard deviation of 0.1 to 0.9 in steps of 0.1, generating a total of 270 degraded spectra.

For each of these degraded spectra, we measure the new S/N and the LOS velocity, with an starting initial guess velocity of 200 \km. 
In Fig.~\ref{fig:noise}, we show the S/N vs the pPXF measured velocity for each of these spectra.
The vertical blue line indicates the actual velocity of the star. It is evident that the scatter in the measured velocities widens as the S/N decreases. 
The scatter in the measured velocities give us a sense of the uncertainty of the measured velocity for a spectra of a given S/N level.
Notably, the uncertainty in measured velocity experiences a stark increase below S/N<3.
From our sample of RGB stars, only one exhibits a S/N>3, yielding a velocity measurement consistent with NGC 4945 (see below in Section~\ref{sec:individual_stars}). Stars with S/N>4 correspond to magnitudes above the TRGB as delineated by the vertical line in Fig.~\ref{fig:snr}. Given our aim to differentiate MW stars we decided that a S/N$=4$ is the minimum S/N that we can use to obtain reliable velocity measurements for individual stars. 
Consequently, we can accurately determine velocities for stars brighter than the TRGB. This capability enables us to discern between most MW stars and potentially bright NGC 4945 stars.
In the outer disk field, we do not have high enough S/N to be able to measuring individual star velocities (see bottom panel in Fig.~\ref{fig:snr}). 

For the second test, i.e. how our velocity measurements depend on the initial guess velocities input in pPXF, we input 6 initial velocities: 0, 200, 400, 600 and 800 \km. We measured velocities using the full wavelength range (4800 to 8800 \angstrom) in stars brighter than the TRGB, both from the GHOSTS catalog and those detected by Gaia which are not in the GHOSTS catalog, for a total of 63 stars. For stars with magnitudes F814W < 22.6, and S/N from 9 to 40, there is no variation in the velocity distribution measured when using different initial velocities. 

For the 24 stars with magnitudes between F814W = 22.6 and the TRGB, the velocities obtained are more sensitive to the initial velocity guess, often tending to result in velocities close to the initial guess. Within this magnitude range, we anticipate the possibility of identifying at least one star belonging to NGC 4945, given the number of RGB stars in the halo field (see \citealt{harmsen2023}). 
Among these stars, three emerge as strong candidates for belonging to NGC 4945. Two of these candidates consistently exhibit similar velocities across all tested starting points, while the third deviates significantly only when an initial guess of 0 km/s is used, registering velocities exceeding 400 km/s with other starting points. Additionally, several stars display velocities closely aligned with the chosen starting guess velocity.
This is expected since pPXF uses a local minimization algorithm to converge in a local minimum. Given that in this magnitude range the individual S/N per star is lower, reaching the limit that gives us reliable velocities, this will cause the velocities obtained to be more dependent on the initial guess velocity. 

\begin{figure}
    \centering
    \includegraphics[width=\columnwidth]{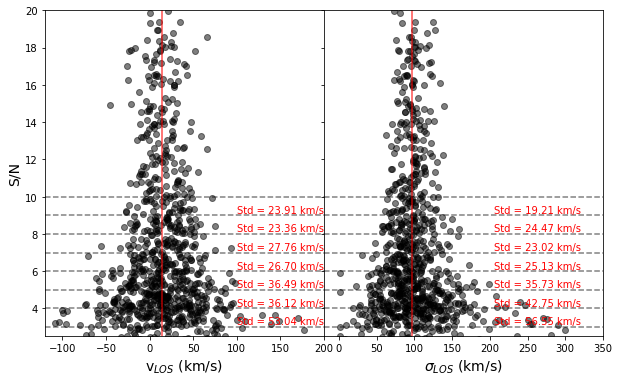}
    \caption{Scatter in the measurements of the velocity and velocity dispersion for stack spectra, resembling spectra of stacked RGB stars, as a function of the S/N. The red vertical lines indicate the median values of velocity and velocity dispersion, 14 and 96 \km respectively. We also report the standard deviation at various levels of S/N, which decreases as the S/N increases.}
    \label{fig:sigma_test}
\end{figure}

In addition to testing the reliability of individual star measurements, we need to test the minimum S/N required to obtain a reliable velocity and velocity dispersion for a stacked spectrum. To do this test, we use a star from the MUSE spectral library \citep{ivanov2019} (S/N=79 and a velocity of 19 \km) as a template to simulate a realistic stack of RGB stars spectrum. We generate a Gaussian distribution comprising 53 velocities centered at 19 \km with a width of 100 \km, representative of a halo field, and applied the corresponding shifts to the original spectrum to mimic the different velocities generated.
We use the S/N and magnitudes of the RGB stars in our MUSE halo field (see Figure~\ref{fig:snr}) to emulate a stack of stars in the RGB that have different S/N according to their magnitudes by introducing noise in their respective spectra. 
Then, we combine the 53 spectra  weighted by their respective S/N values. This combined spectrum serves as the basis for testing velocity values in accordance with varying S/N levels.

We degrade the quality of this combined spectrum by adding random Gaussian noise ranging from 0.1 to 0.9 of width in intervals of 0.05. For each noise level, we generate 50 perturbed spectra and we measure their LOS velocity and velocity dispersion for progressively lower S/N values. Fig.~\ref{fig:sigma_test} shows the scatter in the velocity and the velocity dispersion as the S/N decreases in these spectra. At a given S/N, it is more reliable to measure the mean velocity of a stack than the second-order velocity dispersion.  From this test, we conclude that we can measure the LOS velocity  and the velocity dispersion of the stacked spectrum with sufficient accuracy for a spectrum with S/N>6. 
The uncertainty in the LOS measured velocities, both for the individual stars and for the stacked spectrum, and the uncertainties in the velocity dispersion of the stacked spectra were calculated running 1000 Monte Carlo bootstrapping simulations \citep{cappellari2023}.

\section{Results}\label{sec:results}

We start this section by presenting the results we obtain from measuring the velocities of individual stars in the halo field. We then  show the results obtained from co-adding the spectra of individual stars, selected as in Section~\ref{sec:rgb_stacked}, to obtain the mean velocity and velocity dispersion of the halo and outer disk field.

\subsection{Individual star velocity measurements in the halo field} \label{sec:individual_stars}

\begin{figure}
    \centering
    \includegraphics[width=\columnwidth]{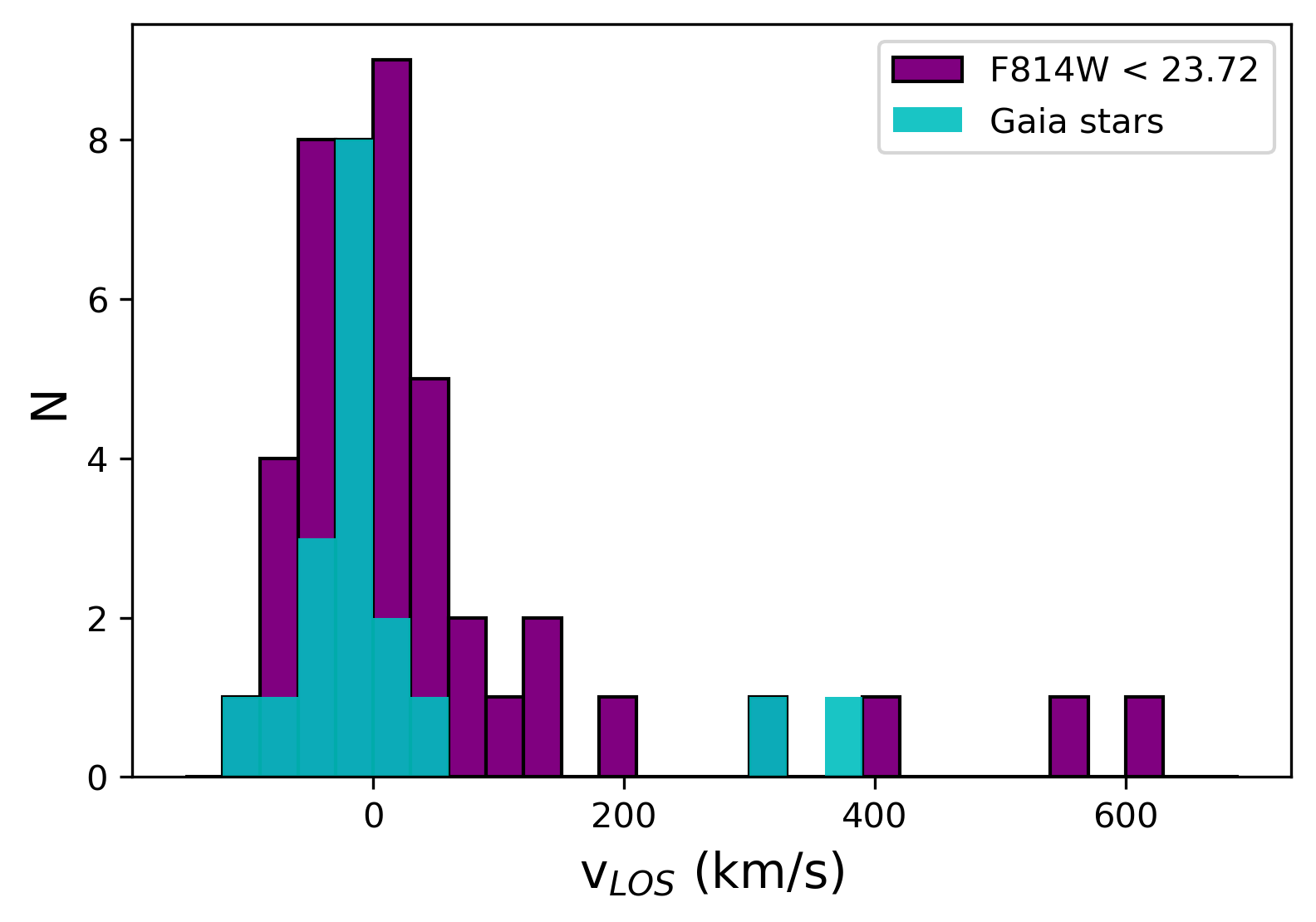}
    \caption{Velocity distribution of the halo field stars brighter than the TRGB (F814W = 23.72). Stars in the GHOSTS catalog are shown in purple while stars detected by Gaia are shown in Cyan.}
    \label{fig:vel_bright}
\end{figure}

\begin{figure*}
    \centering
    \includegraphics[width=\hsize]{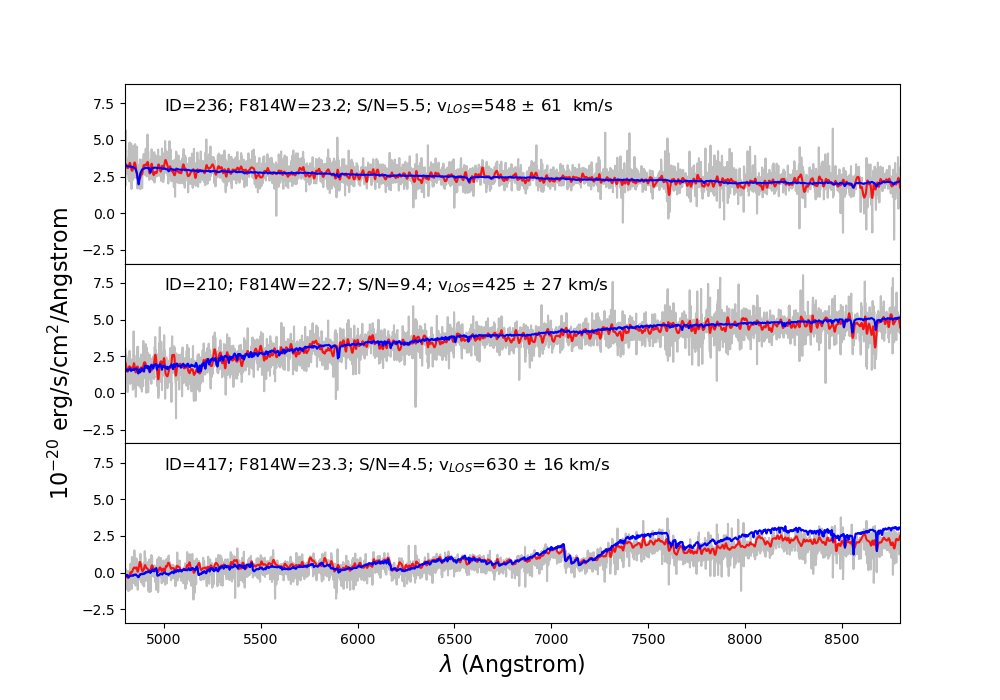}
    \caption{Individual spectra of the three candidate stars (gray) that belong to NGC 4945, according to our velocity criteria to discriminate foreground MW stars, with their respective best fit obtained with pPXF (blue). In red we show the spectra smoothed by a Gaussian kernel of 3 pixels weighted by the inverse variance of the sky spectrum.}
    \label{fig:agb_cand}
\end{figure*}

\begin{table}
	\centering
	\caption{Positions (RA and Dec), magnitude, color, S/N and LOS velocities for the individual 3 stars brighter than the TRGB which are candidates to belong to NGC 4945 based on their measured LOS velocities (see Section~\ref{sec:individual_stars}).}
	\label{tab:indv_stars}
     \setlength{\tabcolsep}{3pt}  
	\begin{tabular}{lcccccc} 
		\hline\hline
            \noalign{\smallskip}
            ID & 236 & 210 & 417\\
            \hline
		RA (h:m:s) & 13:07:40.5 & 13:07:43.9 & 13:07:41.1 \\
            Dec (d:m:s) & -49:03:30.0 & -49:02:46.9 & -49:02:41.3 \\
            F814W (mag) & 23.2 & 22.7 & 23.3 \\ 
            (F606W-F814W) & 0.4 & 1.0 & 2.1 \\
            S/N &  5.5 & 9.4 & 4.5\\
            v$_{LOS}$ (\km)& 548$\pm$61 & 425$\pm$27 & 630$\pm$16  \\		
		\hline
            \noalign{\smallskip}
	\end{tabular}
\end{table}

To decrease the contamination of MW dwarf stars to the co-added spectra of RGB stars in the halo field, we want to kinematically identify the stars belonging to NGC 4945. As shown in Fig.~\ref{fig:snr}, the spectra of stars in the MUSE halo field that are brighter than the TRGB have S/N$\geq$4 per resolution element, thus we can measure their individual velocities reliably (see Fig.~\ref{fig:noise}).  We measure the LOS velocity of these 45 stars in the halo field brighter than the TRGB in the GHOSTS catalog with pPXF, as described in Sec.~\ref{sec:ppxf} and show the distribution of their LOS velocities in Fig.~\ref{fig:vel_bright}. We find that 40 stars have LOS velocities between $-$100 and 200 \km, one has a LOS velocity of 202 \km, one has a velocity of 329 \km and finally there are 3 stars with v>400 \km (Fig.~\ref{fig:vel_bright}). In addition to the stars resolved by GHOSTS, we also measure the velocities of 18 bright high S/N stars detected by Gaia in our MUSE halo FoV. Their LOS velocities measured by pPXF give us a sense of the distribution of velocities of the MW foreground stars. We find that 16 of them have velocities between $-$150 and 50 \km while two have higher velocities, i.e, 302 and 366 \km respectively (see Fig.~\ref{fig:vel_bright}). While most MW foreground stars have velocities distributed around 0 \km, a small fraction can have extremely large velocities. The measured LOS velocities of the Gaia stars supports the assumption that the vast majority of stars with magnitudes brighter than the TRGB are indeed foreground MW stars, consistent with our expectations of this low-latitude FoV. Since the systemic heliocentric velocity of NGC 4945 is 563 \km, and the Gaia stars have all velocities lower than 400 \km, we define an upper limit of 400 \km to discriminate between MW and NGC 4945 stars. Thus, we consider stars with heliocentric LOS velocities greater than 400 \km to be candidates to being NGC 4945 stars.

We found three stars in the GHOSTS catalog brighter than the TRGB that have heliocentric LOS velocities greater than 400 \km, \camila{they are highlighted with small green, blue and red boxes in Fig.~\ref{fig:cmds}}. We note that their magnitudes lie within the magnitude range where our tests showed that measured velocities depend on the initial guess velocity input to pPXF (see previous Section \ref{sec:ppxf}). However, we applied the different tests described in Sec.~\ref{sec:ppxf} to these three stars in particular and found that these three stars have very stable LOS velocities, regardless of the initial guess velocities used (from 0 to 800 \km spaced by 200 \km). Based on kinematics alone, these stars are strong candidates to belong to NGC 4945; their main properties are listed in Table~\ref{tab:indv_stars} and they are highlighted as blue squares in the CMD presented in Figure~\ref{fig:cmds} (left panel). 

Figure~\ref{fig:agb_cand} shows the spectra of these three stars and the best fit obtained by pPXF. 
The reddest star (ID=417) has a LOS velocity of 630$\pm$16 \km and falls within the AGB selected-region of the CMD. Its spectrum shows molecular bands indicative of a low temperature star and is of spectral type M. On the other hand, the bluest star (ID=236) has a LOS velocity of 548$\pm$61 \km, similar to the systemic velocity of NGC 4945. The star situated between them (ID=210) has a LOS velocity of 425$\pm$27 \km, its spectrum shows a slight increase in flux toward redder wavelengths and the CaT lines are distinguishable. The spectra of these last two stars lack the molecular bands seen in the reddest star, consistent with their bluer color (i.e., higher temperature). The bluest star (ID=236) shows evidence of $\mathrm{H}\beta$ line as well as HeI, typical of B stars. Based on their positions in the CMD, these two stars are more likely a Blue Helium burning (BHeB) and a Red Helium burning (RHeB), respectively.
Nevertheless, we cannot rule out the possibility that these could be very high-velocity foreground MW stars. However, this possibility is  low, considering the low number of MW stars with v>400 \km.  Utilizing Gaia DR3 data for stars with G<15, we find that there are only two out of 5000 MW stars with Gaia measured velocities larger than 400 \km ( their velocities are 431 \km and 605 \km), i.e a fraction of 0.0004, in a $2 \times 2$ square degree in Galactic latitude and longitude at the position where our MUSE halo field is located. Considering that in our MUSE FoV of 1 arcmin$^2$ there are 18 Gaia stars, this implies a probability of having 0.0072 MW stars with v>400 \km in the halo field. This is reflected in our measurements as none of the Gaia stars within our Muse FoV have velocities exceeding 400 \km.

We will discuss the implications of these three stars belonging to NGC 4945 in the Discussion section \ref{sec:discussion}, taking into consideration the results of the mean halo velocity and outer disk field, which we present below. 

\subsection{Velocity of the halo field from stacked RGB stars} \label{sec:halo_stacked}

\begin{figure*}
\centering
  \includegraphics[width=1.8\columnwidth]{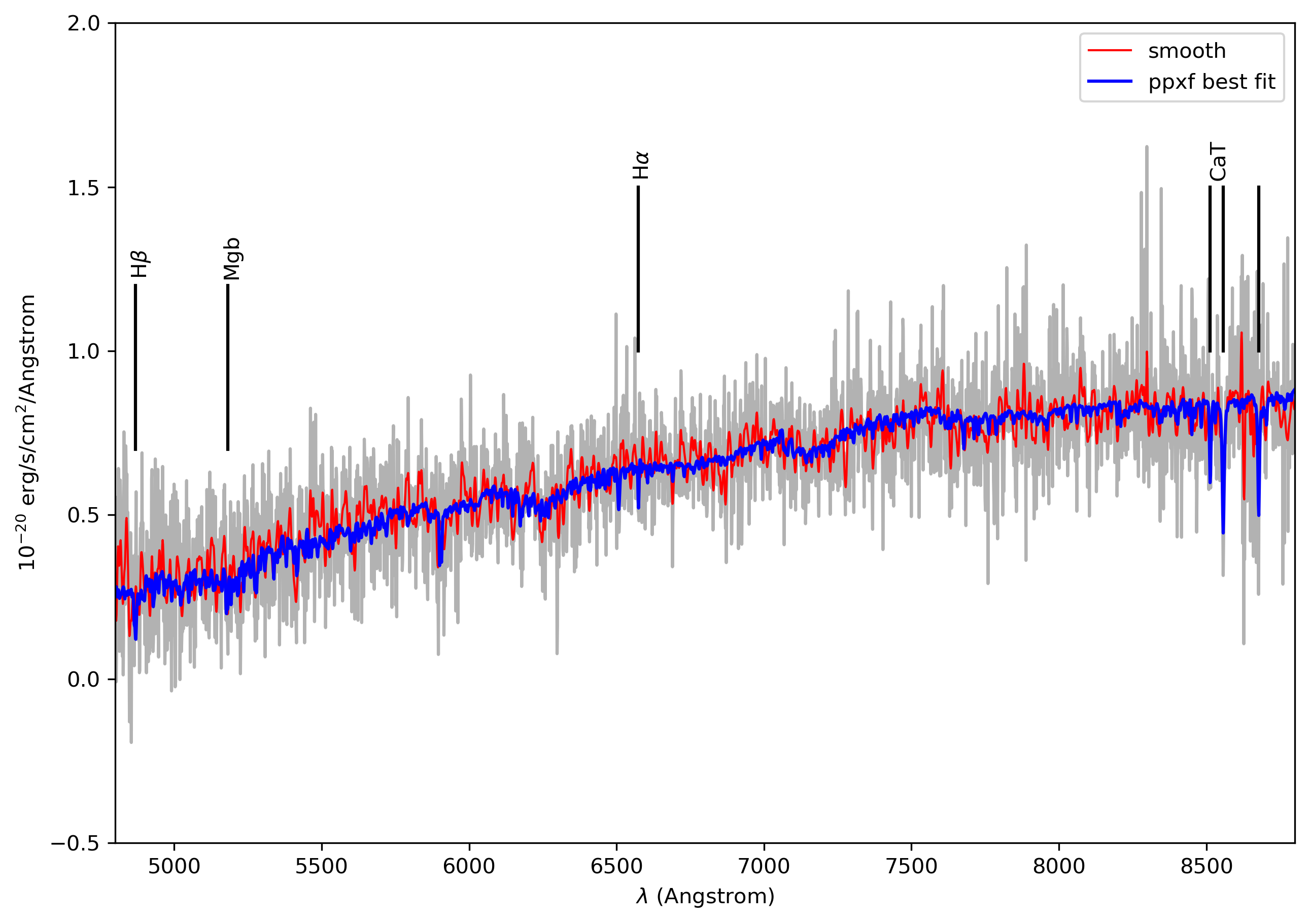} 
    \includegraphics[width=1.8\columnwidth]{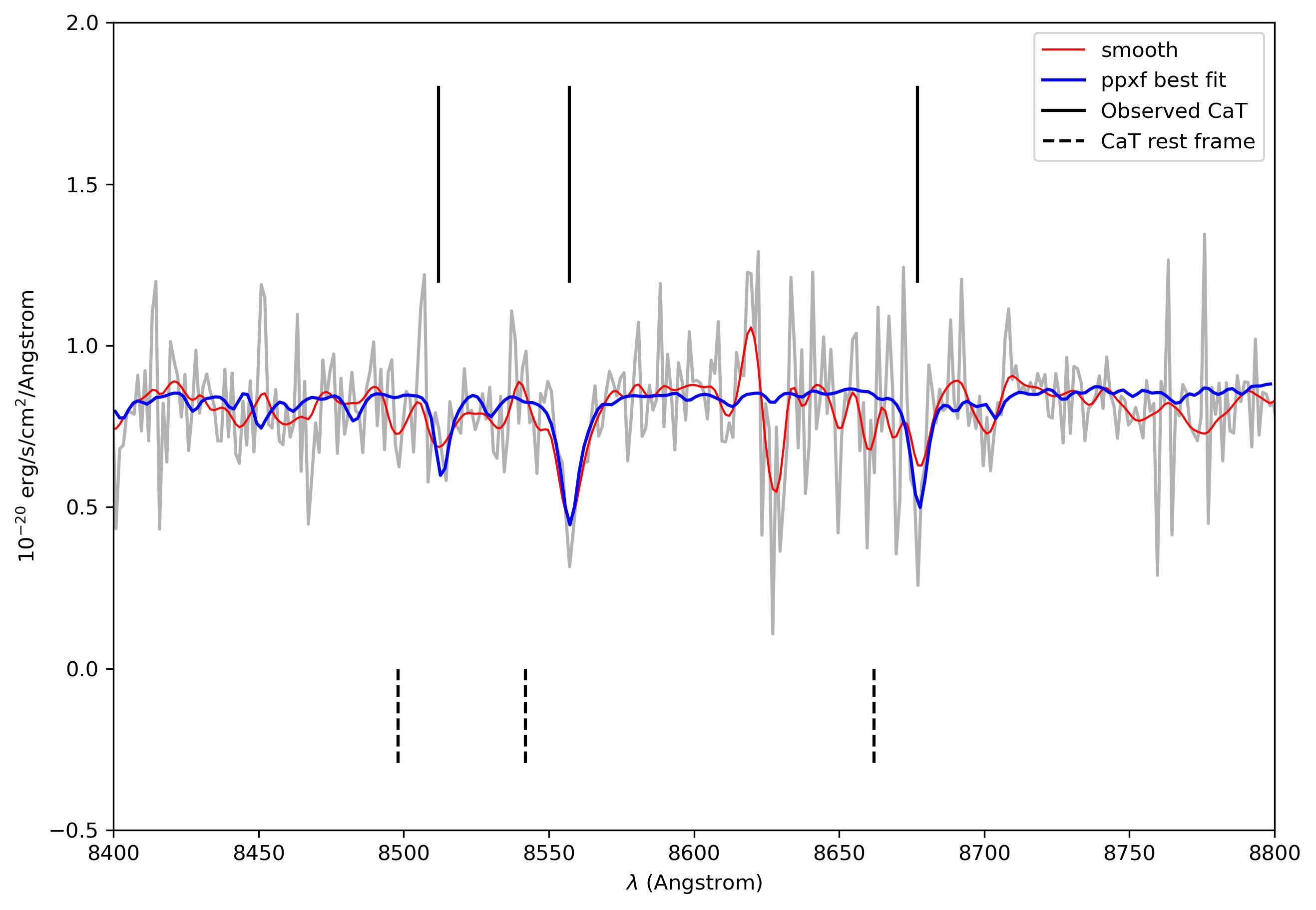}
  \caption{The stacked halo spectrum: Co-added spectrum of 53 halo RGB stars with F814W magnitude brighter than 25.72. Red: Co-add spectrum smoothed by a Gaussian kernel of 2 pixels weighted by the inverse variance of the sky spectrum. Blue: The best fit from pPXF. In the upper panel we show the full spectrum wavelength, 4800-8800 \angstrom, used to measure velocity with pPXF, in the bottom panel we show a zoom in the CaT region and mark with a black dashed lines the position of these lines: 8498,8542 and 8662 \angstrom in the restframe and in black solid lines the position of the observed CaT lines.}
  \label{fig:halo_stacked}
\end{figure*}

We now measure the velocity of our stacked RGB halo stars using pPXF. In brief, we co-added the spectra of RGB stars (that were not significantly contaminated by bright neighbors) extending to 2 magnitudes below the TRGB: we generated 4 stacks reaching down to four limiting magnitudes, including progressively more fainter stars with lower S/N. In Table ~\ref{tab:combine_halo}, we list the magnitude range, the number of coadded RGB stars, the achieved S/N, and the heliocentric LOS velocities and velocity dispersions measured by pPFX with their corresponding uncertainties, for each stacked spectrum.
The four stacked spectra exhibit LOS velocities ranging from 515 to 537 \km.
The S/N in all four stacks is consistently above 8. However, the increase in S/N is not linear with increasing limiting magnitude (i.e. the increasing number of stars), implying that we might be adding more noise than signal. However, considering that the variations in the obtained LOS velocity are not very pronounced, and the S/N values are also relatively stable across the stacks, we adopt the stack N$^\circ$4 as the fiducial stack. This decision is based on it having the highest number of stars. The uncertainties reported are calculated from the Monte Carlo bootstrapping method presented in Section~\ref{sec:ppxf}.

The top panel of Fig.~\ref{fig:halo_stacked} displays the halo spectrum in gray. The blue line represents the best-fitting spectrum obtained using pPXF, while the red line illustrates the smoothed spectrum. The smoothing process employs a Gaussian Kernel with a sigma of 2 pixels and is weighted by the inverse variance of the sky spectrum. The bottom panel of this Figure is a zoom in the CaT region in which we can see that the two strongest CaT lines are very clear in the spectrum.

We obtain a mean LOS velocity of 519 $\pm$ 12 \km and a velocity dispersion of 42$\pm$22 \km for the halo field. We emphasize that this is the \textit{first ever measurement of the kinematics of a diffuse stellar halo outside the Local Group obtained from its resolved stars}. This halo field LOS velocity measurement is within $\sim$ 40 \km from the systemic velocity of 563 \km for NGC 4945 \citep{koribalski2004}. Considering that NGC 4945 is edge-on, its measured LOS velocity on the major axis subtracted from its systemic velocity can be considered its rotational velocity. This is $-44$ \km , meaning that at this distance the halo is counter-rotating. This result suggests a clear evidence of an accreted stellar halo at that distance. We will discuss the consequences of this result in Section \ref{sec:discussion}.

\camila{We note that our velocity dispersion (42$\pm$22 \km) is below the MUSE velocity resolution ($\sim$50 \km). To assess the ability of pPXF to recover such velocity dispersion, we performed the same test as shown in Fig.~\ref{fig:sigma_test} but using a Gaussian distribution with a width of 40 \km (see Sec.~\ref{ap:sigma_test}); i.e. we generate a mock stack spectrum of RGB stars with a velocity dispersion of 40 \km. We found that pPXF successfully recovers the input velocity dispersion of 40 \km in that test, providing a median value of 39 \km . This is shown in Sec.~\ref{ap:sigma_test}, see Fig.~\ref{fig:sigma_width40}. This demonstrates the code's ability to accurately retrieve low velocity dispersion values, as these are extracted from the best fit the code produces to the real spectrum. However, given that our measured velocity dispersion is lower than the MUSE resolution, we cannot provide a more precise value than 50 \km, so our result of 42$\pm$22 \km will be the upper limit of the actual velocity dispersion in our halo field.}

Since there are three stars brighter than the TRGB that are possibly NGC 4945 stars (see previous section), we examine the resulting velocity when adding them to the combined spectrum. We generated a new stacked spectrum by incorporating these three stars into stack N° 4, resulting in a final S/N of 14.5, and a velocity and velocity dispersion of 510 $\pm$ 13 \km and 71 $\pm$ 18 \km, respectively.

\begin{table}
	\centering
	\caption{Halo co-add stacked of RGB stars at different magnitude levels as described in Sec.~\ref{sec:halo_stacked}. The listed uncertainties are measured by running 1000 Monte Carlo bootstrapping simulations.}
	\label{tab:combine_halo}
	\begin{tabular}{clcccc} 
		\hline\hline
             \noalign{\smallskip}
		Stack & F814W & N & S/N & v$_{\rm LOS}$ & $\sigma$ \\
		 N$^{\circ}$& & $\#$ &  & (\km) & (\km)\\
		\hline
            \noalign{\smallskip}
		1 & 23.72-24.72  & 14 & 8.9 & 537 $\pm$ 16 & 48$\pm$27\\
		2 & 23.72-25.00 & 28 & 9.8 & 515 $\pm$ 14 & 53$\pm$22\\
		3 & 23.72-25.50 & 43 & 9.7 & 517 $\pm$ 13 & 46$\pm$23 \\
		4 & 23.72-25.72 & 53 & 9.4 & 519 $\pm$ 12 & 42$\pm$22\\
		\hline
	\end{tabular}
\end{table}

\subsection{Velocity of the outer disk field from the stacked spectrum}\label{sec:outer_disk}

The low S/N of individual stars in the outer disk field  prevents us from kinematically distinguishing MW foreground stars from stars proper to NGC 4945. However, this is not a problem for the outer disk field since the number of MW foreground stars is negligible compared to the number of RGB and, especially, AGB stars in NGC 4945 (see Fig.~\ref{fig:cmds}). This contamination can be roughly estimated from the Trilegal simulation: 10 predicted MW foreground stars vs. 91 stars that we have in total in the AGB box region in Fig.~\ref{fig:cmds}, which accounts for a contamination of less than 20\%. Along with a clear and well populated RGB, we can discern a clear AGB population in the disk field. Considering that we have 1 or 2 AGB star candidates in the halo field where there are 68 RGB stars down to 2 magnitudes below TRGB, we estimate that there should be, in proportion, at least 53 AGB stars in the outer disk field, given that there are 1835 RGB stars within the RGB box. This estimate is consistent with what we see in the CMD (Fig.~\ref{fig:cmds}).  Nevertheless, to avoid MW star contamination as much as possible, we decide to be conservative in the color of the AGB region considered to do the stack, with a red limit of 2.5, since MW contamination is higher in the redder part of the AGB (see \citealt{monachesi2016a}). 

As mentioned in Section \ref{sec:extraction}, to analyze this field we first divide it into 4 quadrants: The upper and lower left (Q1 and Q3 respectively) and the upper and lower right (Q2 and Q4), as shown in Fig.\ref{fig:FoV}. This is because the crowding of the field increases significantly from Q1 to Q4, thus the background in each of these quadrants is also different.  In each of the quadrants we select all the RGB/AGB stars free of contamination from neighbor stars (as explained in Sec.~\ref{sec:contamination}) and combine their spectra.  We also generate, as in the halo field, four distinct stacked spectra, each including RGB stars with fainter magnitudes, reaching down to 2 mag below the TRGB. All 70 AGB stars are included in each of the stacks.  Table~\ref{tab:combine_disk} displays the number of RGB stars used in each combined spectrum, the S/N of the combined spectrum with both RGB and AGB stars (values within parentheses represent the results considering only the RGB stars), and the measured LOS velocities and velocity dispersion values. We highlight that across all stacks, the obtained LOS velocity values exhibit stability, with differences of less than 6 \km between them. Additionally, we find that, even though the shallow outer disk field has a much lower exposure time, the higher density of RGB and AGB stars belonging to NGC 4945 compared to the halo field compensates in a way that it is possible to obtain a final co-added spectrum with a much higher S/N ($\sim$16) compared to the latter ($\sim$9).

\begin{figure*}
\centering
\includegraphics[width=1.8\columnwidth]{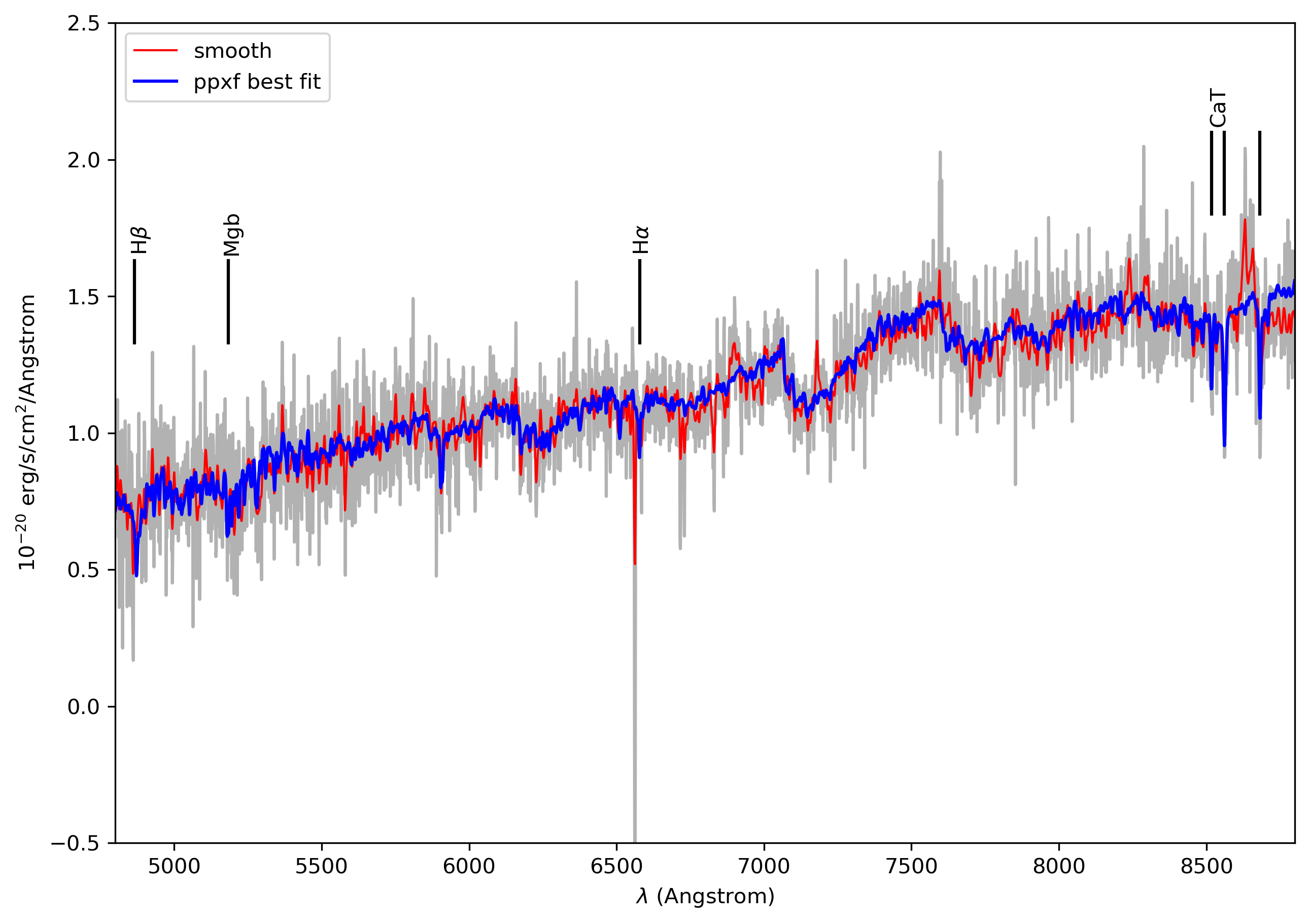} 
    \includegraphics[width=1.8\columnwidth]{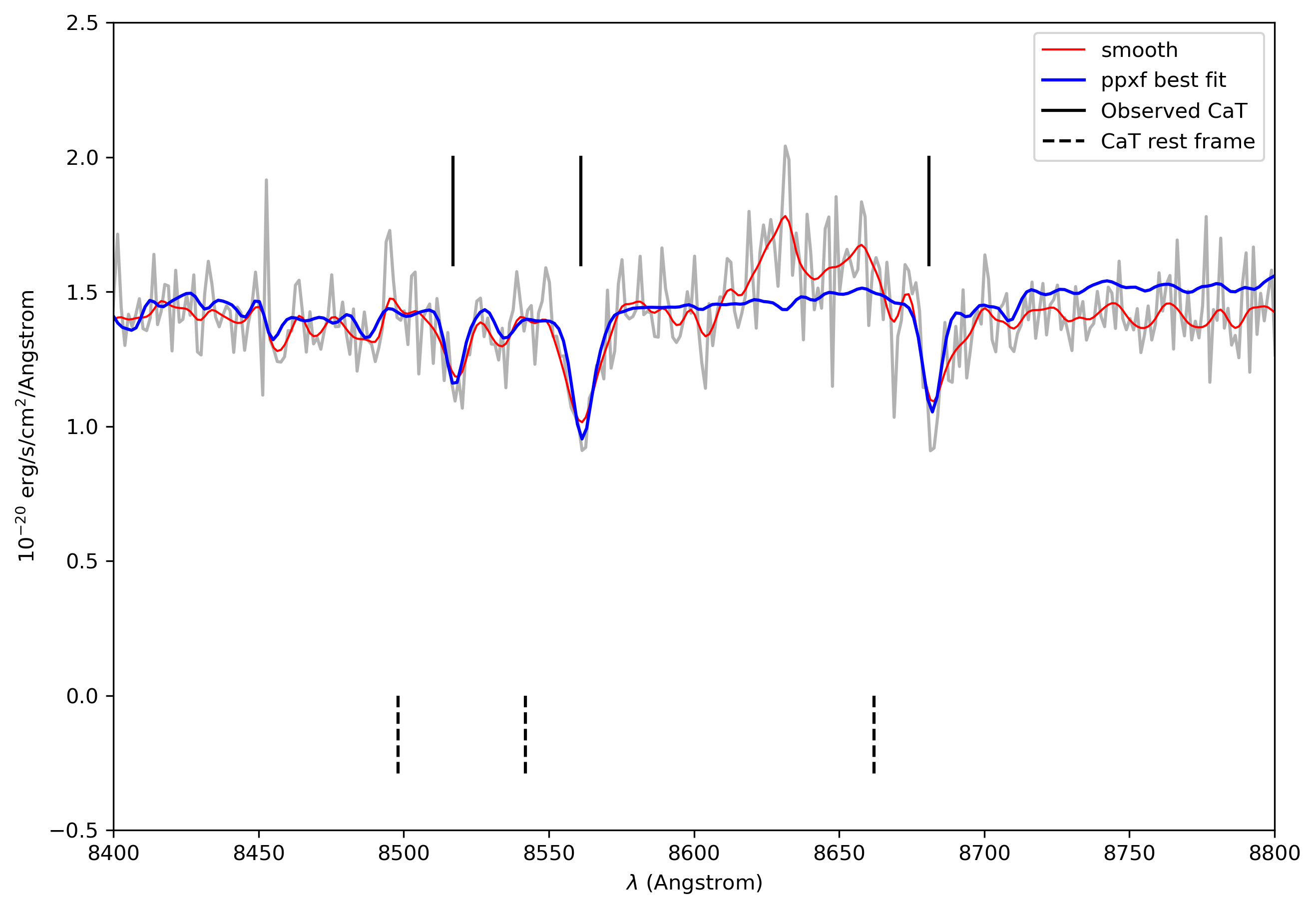} 
      \caption{The stacked outer disk spectrum: Grey represents the mean co-added spectrum of AGB plus RGB stars with F814W magnitude brighter than 25.72. Red: Co-added spectrum smoothed by a Gaussian kernel of 2 pixels weighted by the inverse variance of the sky spectrum. Blue: The best fit from pPXF. In the upper panel we show the full spectrum wavelength, 4800-8800 \angstrom, used to measure velocity with pPXF, in the bottom panel we show a zoom in the CaT region and mark with a black dashed lines the position of these lines: 8498,8542 and 8662 \angstrom in the restframe and in black solid lines the position of the observed CaT lines.}
  \label{fig:disk_stacked}
\end{figure*}

For consistency, to select the spectrum that best represents the outer disk, we base our choice on the one with the highest number of stars in its stack, as we did for the halo. Stack N$^{\circ}$4 meets this requirement, having a S/N=16.7, so this will be used as representative for our analysis.
The top panel of Fig.~\ref{fig:disk_stacked} shows the combined outer disk field spectrum of the stack N$^\circ$4 of Table~\ref{tab:combine_disk}. We also show the smoothed spectrum in red using a Gaussian kernel of 2-pixel weighted by the inverse variance of the sky spectrum. Finally we show the best fit calculated by pPXF as blue line. The bottom panel shows the zoom-in of this spectrum in the CaT region, where the three lines are clearly detected. We obtain a mean heliocentric LOS velocity of 673 $\pm$ 11 \km for the outer disk field, which is in good agreement with the average velocity obtained from the mean HI gas velocity of the disk \footnote{\url{https://www.atnf.csiro.au/research/LVHIS/data/LVHIS043.info.html}} \citep[$\sim$700 \km,][]{koribalski2018} near the position of our target. Given the high S/N of this final stack, and the results of the tests performed in Section~\ref{sec:ppxf}, we can trust the velocity dispersion measurement obtained for this field, which is 73$\pm$14 \km.

As a consistency check, we also derived the LOS velocity of the outer disk field using only RGB stars. We found that the combined spectrum using only RGB stars has a lower S/N, as expected (see Table~\ref{tab:combine_disk}), however the second and third lines of the CaT are still distinguishable and the obtained LOS velocity is 677$\pm$11 \km, consistent with that obtained when also the AGB stars are combined.

\begin{table}
	\centering
	\caption{Outer disk co-add stacked of 70 AGB plus RGB stars at different magnitude levels. The values of S/N, velocity and velocity dispersion are obtained by considering the co-add stacking of RGB and AGB stars. The values within parentheses are the S/N considering only the stack of RGB stars. The listed uncertainties are measured by running 1000 Monte Carlo bootstrapping simulations.}
	\label{tab:combine_disk}
    \setlength{\tabcolsep}{3pt}
	\begin{tabular}{clcccc} 
		\hline\hline
            \noalign{\smallskip}
		Stack & F814W & RGB & S/N & v$_{\rm LOS}$ & $\sigma$ \\
		 N$^{\circ}$&(mag) & $\#$ &  & (\km) & (\km)\\
		\hline
            \noalign{\smallskip}
		1 & 23.72-24.72  & 427 & 16.2 (14.9) & 667$\pm$11 & 76$\pm$16 \\
		2 & 23.72-25.00  & 622 & 15.6 (14.6)& 670$\pm$12 & 79$\pm$15\\
		3 & 23.72-25.50  & 989 & 16.3 (14.7) & 671$\pm$11 & 76$\pm$15 \\
		4 & 23.72-25.72  & 1122 & 16.7 (14.8) & 673 $\pm$11 & 73$\pm$14\\
		\hline
	\end{tabular}
\end{table}

\section{Discussion} \label{sec:discussion}

In this work, we measured the median LOS heliocentric velocities of two fields along the major axis of the edge-on galaxy NGC 4945, positioned at 12.2kpc (outer disk) and 34.6 kpc (halo) from the center of NGC 4945, by stacking their RGB and AGB stars. Since NGC4945 is edge-on, its measured velocities along the major axis can be regarded as its rotational velocity. Our findings reveal that the field at 12.2kpc has a LOS velocity of 673 $\pm$ 11 \km, in agreement with the velocity of the gas in a nearby position, which is $\sim$700 \km \citep{koribalski2018}. This suggests that the stars in the outer disk field are rotating and are an extension of the disk of the galaxy. On the other hand, the field at 34.6 kpc, representing the stellar halo, displays a LOS velocity of 519 $\pm$ 12\km, which is $\sim$40 \km lower but consistent with the systemic velocity of NGC 4945 \citep[563 \km,][]{koribalski2004}. Consequently, at this distance, the stellar halo of NGC~4945 does not show signatures of strong rotation and it appears to be counter-rotating, implying that this is an accreted halo. Within the halo field, we also identified three stars, brighter than the TRGB, with LOS velocities exceeding 400 \km. These stars are potential candidates for belonging to NGC 4945, and if so these are likely a BHeB, an RHeB, and an AGB star, based on their positions in the CMD and their spectra. 

In this section, we further discuss the limitations of our employed technique and possibilities of using this technique to measure kinematics in other galaxies. We then present a global picture of NGC 4945 stellar halo, combining our results within the existing knowledge of the stellar halo of NGC 4945 from previous work and including a discussion of the three bright stars possibly belonging to NGC 4945 stellar halo. We contextualize our results of the stellar halo of NGC 4945 in comparison with the results obtained for the stellar halos of MW and M31 -- the only other two galaxies for which there are velocity measurements of their halos based on individual RGB stars. To conclude this section, we place everything into a broader perspective of halo formation by comparing our results with the Auriga and TNG50 simulations.

\subsection{Limitations of the technique used to derive our measurements}

This is the first measurement of the velocity of a diffuse stellar halo of a galaxy outside the Local Group using resolved stars, at a field with a SB of $\sim$29.5 mag/arcsec$^2$, much fainter than the 27 mag/arcsec$^2$ reached by \cite{toloba2016}. This offers us a taste of what will be routinely possible with the next-generation Extremely Large Telescope (ELT). However, a number of challenges need to be overcome. First, our ability to measure a LOS velocity depends on the S/N of the final stack, which in turn depends upon the number of RGB/AGB stars available to be co-added. In low surface brightness regions (fainter than 28 mag/arcsec$^2$), this becomes extremely challenging. Furthermore, at low Galactic latitudes, one also has to contend with bright MW foreground contamination which further decreases the number of available RGB stars for co-adding. In this work, MW foreground contamination decreases the available number of RGB stars to only 53 in the halo field, resulting in a S/N of 9.4. Note, however, that according to our tests this is sufficient to obtain reliable LOS velocity and velocity dispersion measurements using pPXF (see Section~\ref{sec:ppxf}). 

Secondly, it is very difficult to extract the spectra of stars at very low S/N (<1), especially of stars which are much fainter than the sky background. In this work, we applied a technique akin to forced PSF photometry. Furthermore, a very careful background subtraction is necessary to prevent contamination in the obtained spectra, whether from nearby stars, unresolved sources, or the sky background itself.  
We measured the variation of the sky background across the halo FoV, particularly focusing on the wavelength region between 8400 to 9350 \angstrom, as this is where the greatest differences in spectra are observed. In this region, we found a minimal variation of 55\%, and values reaching up to 130\%. Thus there is a large sky background variation along the FoV, in that wavelength range. In the bluer wavelength range (4800-6800 \angstrom) the sky background variations are similar, ranging from 33\% to 180\%. If we subtract a general sky background to the extracted spectra of the stars, instead of creating a local background spectrum around each star and subsequently subtracted it from the star's spectrum (see Section~\ref{sec:extraction}), we obtained S/N values of around 7 for the combined stack spectrum. Despite the spectra being noisier, we still obtained stable velocity values, ranging from 515 to 526 \km, similar to those shown in the Table~\ref{tab:combine_halo}. This indicates that even with a general background, reliable velocity values can be obtained. This confirms our test results from Fig.~\ref{fig:sigma_test}, which shows that the velocity values can be trusted even at S/N values above 4. Nevertheless, in order to achieve spectra with minimal noise and improve the background subtraction, in the halo field, we created a local background spectrum around each star and subsequently subtracted it from the star's spectrum.

On the other hand, in the outer disk field, creating a local background spectrum around each star was impractical due to the crowding. Instead we adopted a global background, dividing the FoV into four quadrants, each having a mean background spectrum constructed and subtracted from the spectrum extracted for each individual star. While this approach proved adequate for our purposes and give stable velocity measurements, it certainly can be improved by employing smooth 2d splines to fit the background sky spectrum.

\subsection{A global picture of the stellar halo of NGC 4945}

The stellar halo of the MW-like galaxy NGC 4945 has been studied by the GHOSTS survey \citep{monachesi2016a,harmsen2017}. \citet{harmsen2017} derived the density profile of NGC 4945 stellar halo which is found to decrease as a power law, at a similar rate in both axes, with slopes -2.73 and -2.72 measured up to a distance of approximately 40 kpc. They also found an oblate stellar halo shape with projected axis ratio of $c/a \sim 0.52 $ at a galactocentric distance of $\sim$25 kpc. They estimated the total mass of the stellar halo to be approximately 3.5 $\times 10^9\solarmass$, which is about 9$\%$ of the galaxy's total stellar mass.  We highlight that NGC~4945 has a typical stellar halo mass and halo metallicity for a MW-mass galaxy \citep{harmsen2017}, thus we are in this work characterizing kinematically a typical stellar halo for a MW-mass galaxy. 

On the other hand, \citet{monachesi2016a} found a weak RGB color gradient in the stellar halo along the major axis of NGC 4945. This corresponds to a decrease in metallicity [Fe/H] from -0.8 ($\sim$15 kpc) to -0.95 dex ($\sim$45 kpc). It is interesting to note that the metallicity along the major axis of the stellar halo is similar to that found on the minor axis which is rather flat, corresponding to a halo metallicity of [Fe/H] of -0.9 dex \citep{monachesi2016a}. Models suggest that the contribution of the in-situ halo beyond 10 kpc along the minor axis is minimal. The similarity of the metallicities on the minor and major axis of NGC~4945 halo implies that the stellar populations along the major axis beyond 20 kpc resemble the accreted stellar population along the minor axis. The flat metallicity gradient might indicate the contribution of a few (between three and ten) significant satellites at this distance \citep{cooper2010, monachesi2019}.

Our measurement of the LOS velocity of the halo field at 34.6 kpc along the major axis of NGC 4945, 519 $\pm$ 12 \km, given the systemic velocity of 563 \km, shows, for the first time, a counter-rotating halo, thus of an accreted origin. We highlight that NGC 4945 is an edge-on galaxy, thus its measured LOS velocity on the major axis is a good representation of its tangential velocity. This finding is particularly significant given the flattened nature of NGC 4945 halo, challenging the conventional expectation of rotation for such flattened halos (see e.g., the highly rotating accreted stellar halo of M31 \citealt{ibata2005, dsouza&bell2018}). Our result then demonstrates the existence of a flattened, accreted halo counter-rotating. Importantly, we emphasize that reaching such conclusions would have been impossible without the kinematic information we obtained in this work.
Additionally, our measurements reveal a velocity dispersion of 42 $\pm$ 22 \km for NGC~4945 in our MUSE halo field. This velocity dispersion is lower than the typical values of velocity dispersion for halos of $\sim$100 \km (see Section \ref{sec: simulations} for a comparison with cosmological simulations). The lower velocity dispersion could stem from our measurement capturing a segment of a colder halo component, consistent with its flattened shape and the counter-rotation of $\sim$ 40 \km along the major axis. We need more spectroscopic observations into the halo of NGC~4945 to confirm this.

Combining all the results, we can come up with a global observational picture of the stellar halo of NGC 4945. This is a relatively massive ($\sim$3.5$\times 10^9\textup{M}_\odot$) and metal-rich (-0.9 dex) stellar halo, typical for a MW-mass galaxy. Additionally, it has a flattened shape of $c/a \sim 0.52 $ and a rather flat metallicity gradient along the minor axis. Moreover, based on the robust correlation established by \citet{harmsen2017} between the observed stellar mass of the halo and its metallicity at 30 kpc, we gain valuable insights into the properties of the most significant satellite that was accreted by its host. In this context, \citet{bell2017} estimated the most significant satellite's mass based on the stellar halo mass, suggesting that for NGC 4945, the mass of its dominant satellite contributor would be approximately $\sim 1.5 \times 10^9$M$_{\odot}$, similar to the stellar mass of the LMC. In this work we have proved that NGC 4945 halo has a counter-rotation of $\sim-$44 \km at 34.6 kpc along the major axis, which demonstrates its accreted origin.

\subsection{The three candidate bright stars found in the halo field of NGC 4945}

Individual velocity measurements of stars brighter than the TRGB in the halo field suggest that they belong to the halo of NGC 4945. All three stars have a heliocentric LOS velocity greater than 400 \km. Of the three stars (listed in Tab.~\ref{tab:indv_stars}), the reddest star (ID=417) falls within the AGB box (see Fig.~\ref{fig:cmds}). The low number of AGB stars is consistent with the low stellar density of this field and correspondingly, the low number of RGB stars in this field \citep{harmsen2023}. Its heliocentric LOS velocity (630$\pm$16 \km) closely resembles the velocity of the disk of NGC 4945. Besides this AGB star, we also found two stars with bluer colors. These stars closely resemble BHeB and RHeB stars, due to their positions in the CMD \citep{R-S11}, showing evidence of H$\beta$ and HeI lines in their spectra. The presence of Helium burning stars, which have an age between 10-1000 Myr, found at 34.6 kpc from the center of the NGC 4945 is a surprise and cannot be explained by simple radial migration models of galaxy stellar disks. Their relative lower velocities (548$\pm$61 \km and 425$\pm$27 \km ) are closer to the systemic velocity of NGC4945 and do not share the dynamics of the disk. Their velocities then suggest that these stars were probably formed in-situ in its stellar halo. If so we should expect to find HI gas in the halo too. We note that \cite{ianjama2022} found HI gas in the halo of NGC~4945, beyond its optical disk,  however concentrated mainly on the receding side of the galaxy not the approaching side where we have our halo field.

We also do not rule out the possibility that these stars are hyperverlocity MW dwarf stars, given the low latitude of the observed fields ($b=13.3^{\circ}$), which introduces a significant amount of foreground contamination from the MW. 
As discussed in Sec.~\ref{sec:individual_stars}, the probability that these stars are indeed MW foreground is very small ($\approx 0.007$).

\subsection{A comparison with the kinematics of the MW and M31's stellar halos }

The kinematic measurements of the stellar halo of NGC 4945 allow us to directly compare with, and to put into context, the two other MW-like galaxies for which we have kinematic data of their stellar halos, i.e., the MW and M31.  Various surveys have been used to study the halo of the MW, including SDSS \citep{bell2008,deason2011}, APOGEE \citep{mackereth2020}, H3 survey \citep{conroy2019} and Gaia \citep{gaia2016,gaiadr2,gaiaedr3}, among others. Gaia, in particular, has been revolutionary in providing us with precise photometry, positions, velocities, distances, and proper motions for more than 1 billion sources. \citet{deason2017} using a combination of SDSS imaging and Gaia DR1 astrometry measured a small average rotation of $\sim$15 \km in the stellar halo. The radial velocity dispersion profile decreases from 141 \km to 100 \km (in the inner 20 kpc) and then remains almost constant out to a distance of $\sim$70 kpc \citep{bland-harwthorn2016}.
Thanks to Gaia, the detection of Gaia-Enceladus, the fossil of a major progenitor that was accreted 8-10 Gyr, was made possible \citep{helmi2018,Belokurov2018}. These results and the fact that the MW has a small to average stellar halo \citep[$10^9\mathrm{M}_{\odot}$, ][]{deason2019} brings to light that the accretion history of the Galaxy has been rather quiet for the past gigayears compared to similar sized galaxies.

On the other hand, the proximity of M31 provides a unique opportunity to obtain detailed observations of its entire stellar halo, although with less detailed phase-space information than for the MW. Several surveys, such PAndAS \citep[Pan-Andromeda Archeological Survey,][]{ibata2014,mcConnachie2018} and SPLASH \citep[Spectroscopic and Photometric Landscape of Andromeda’s Stellar Halo,][]{gilbert2012,gilbert2018}, have been dedicated to mapping M31's stellar halo using resolved stars. These reveal a massive \citep[1.5 $\pm$ 0.5 $\times$10$^{10}$M$_{\odot}$,][]{ibata2014} and metal-rich \citep[-0.5 dex,][]{gilbert2014}{}{} stellar halo, with a steep metallicity gradient \citep{gilbert2014, escala2020,escala2021}, reflecting the existence of a more massive and recent accretion event, accreted   $\sim$2 Gyr ago, with remnants observable in its giant stellar stream  \citep{dsouza&bell2018,hammer2018}. M31's inner stellar halo shows significant rotation of $\sim$150 \km out to galactocentric distances of 40-70 kpc along the major axis \citep{ibata2005,dey2023}. Additionally,  the velocity dispersion of its stellar halo decreases from 108 \km to $\sim$80-90 \km at a projected distance of $\sim$40-130 kpc from the center \citep{gilbert2018}.

In this work, we are adding a third galaxy to which we have kinematic analysis of its stellar halo with resolved stars. We find that the flattened stellar halo (with projected c/a=0.52) of NGC~4945 is accreted at galactocentric distance of 34.6 kpc, and has a counter-rotational velocity of $\sim-$40 \km. Our measurements also reveal a velocity dispersion of 42 $\pm$ 22 \km,  lower than the typical halo values of $\sim$100 \km. As mentioned before, this velocity dispersion could stem from our measurement capturing a segment of a cold halo component.

\subsection{Insights on stellar halo formation from simulations: A comparison with the Auriga and TNG50} \label{sec: simulations}

\begin{figure}
    \centering
    \includegraphics[width=\columnwidth]{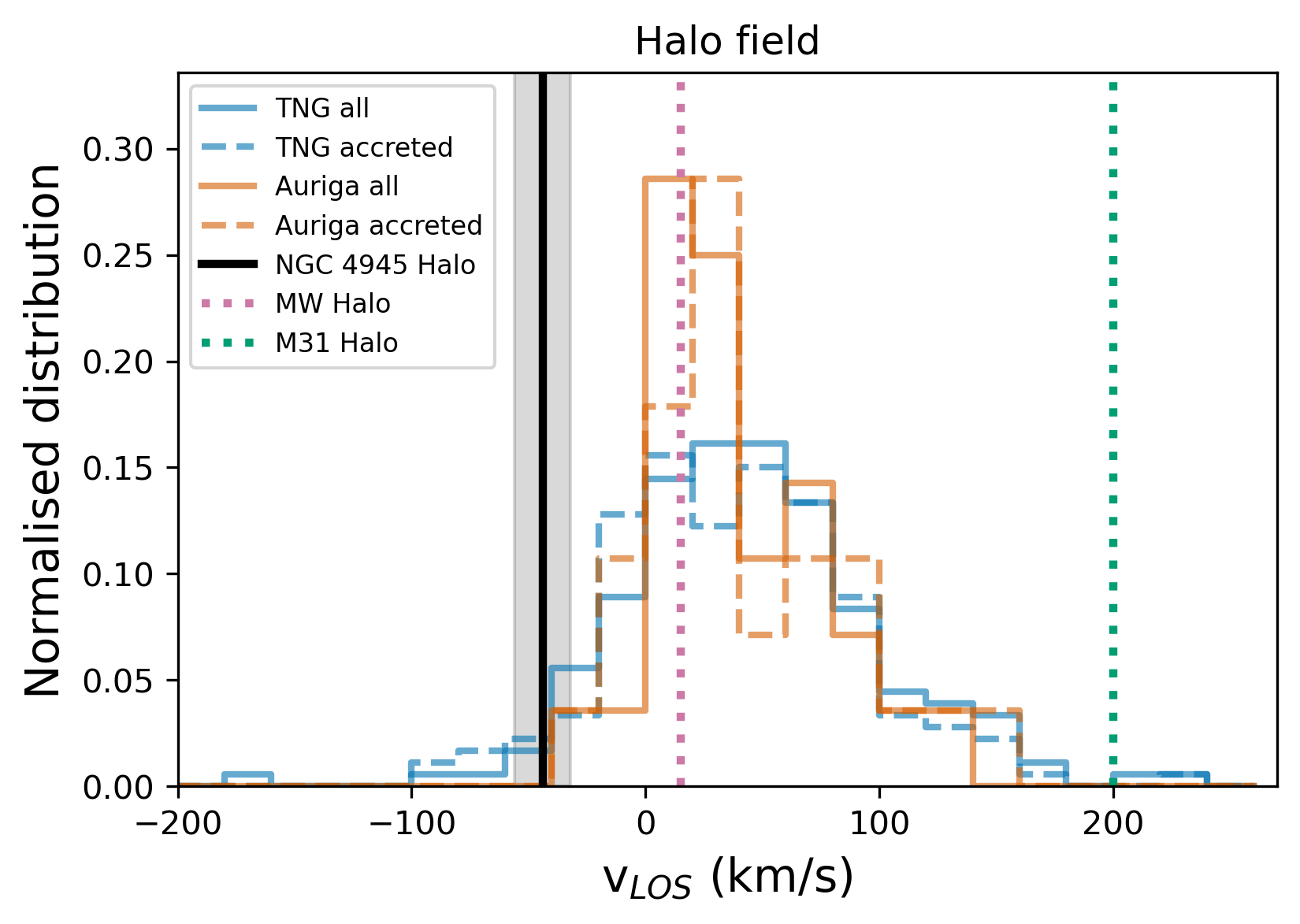}
    \includegraphics[width=\columnwidth]{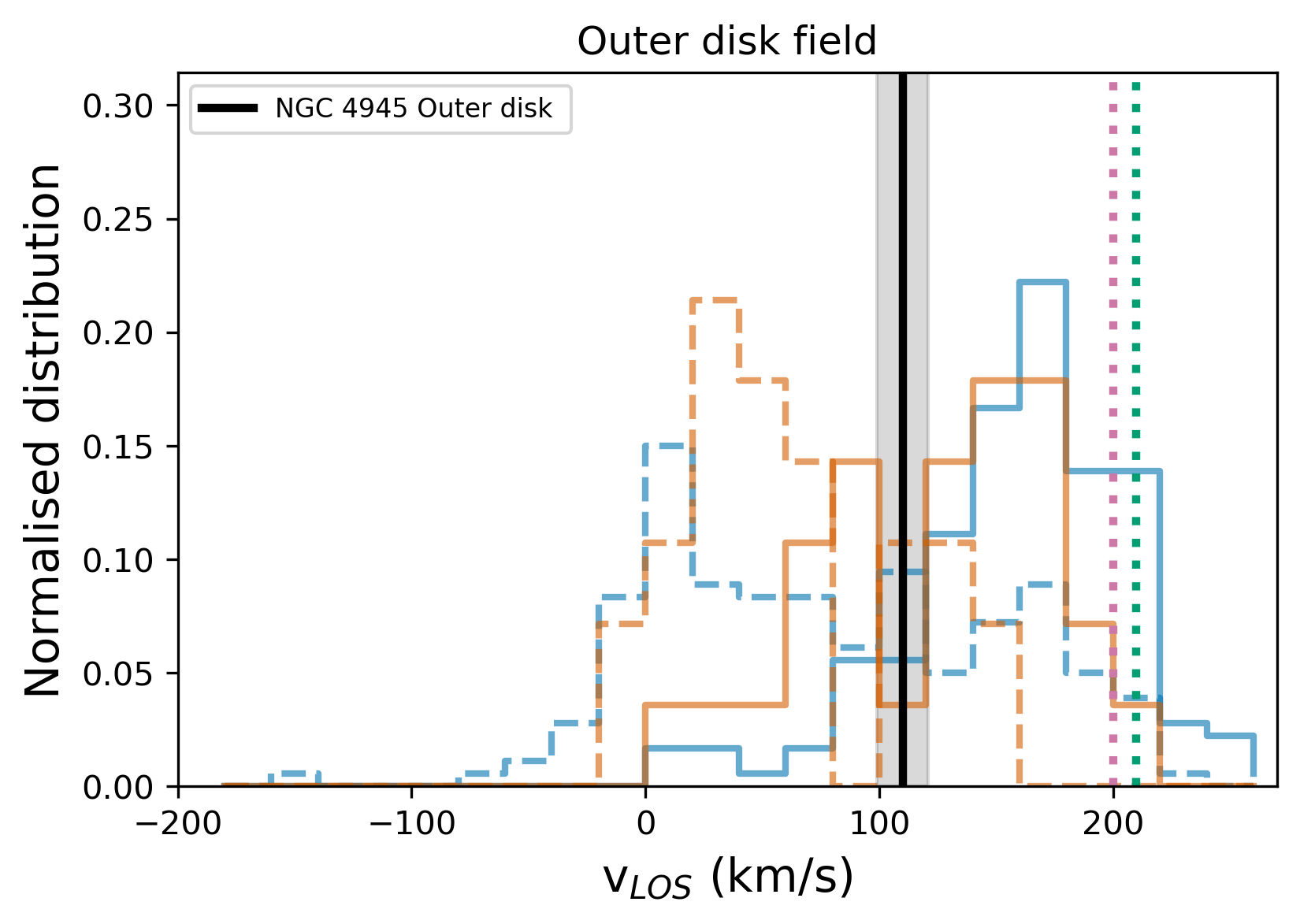}
    \caption{Median  LOS  velocity distribution at galactocentric distances of \camila{1.02R$_{\rm opt}$ (Upper) and 2.9R$_{\rm opt}$} (bottom) kpc along the major axis for MW/M31 like galaxies in an edge-on configuration, in the TNG50 (blue) and Auriga (orange) simulations. Black vertical lines are our velocity measurements for NGC 4945 at those distances, corrected for its systemic velocity. Green and pink dotted vertical lines show the halo rotational velocity of the MW \citep[halo and outer disk,][respectively]{deason2017, bland-harwthorn2016} and M31 \citep[halo and outer disk,][respectively]{ibata2005,zhang2024}. 
    }
    \label{fig:vel_simulation_ropt}
\end{figure}

In this section we compare our results with hydrodynamical cosmological simulations, to gain insights on the kinematics of stellar halos in MW-type galaxies within the $\Lambda$CDM cosmological model and what we can learn about their origin from kinematics. In particular, we use two sets of simulations: Auriga \citep{grand2017} and IllustrisTNG50 \citep{pillepich2019,nelson2019}.

The Auriga project is a set of more than 40 cosmological magneto-hydrodynamical zoom-in simulations of isolated MW-like galaxies. 
These galaxies were chosen based on their dark matter halos selected from the EAGLE project \citep{schaye2015} and fall within the dark matter mass range of $1 \leq M_{200}/10^{12} M_{\odot}\leq 2$, comparable to the mass of the MW. These selected  dark halos were resimulated with the AREPO code \citep{springel2010} at higher resolution.
In this work we use 30 galaxies of resolution level named Level 4 in \citet{grand2017}, i.e. with a baryonic mass resolution of $\sim 5 \times 10^4 \textup{M}_{\odot}$ and a dark matter mass resolution of $\sim 3 \times 10^5 \textup{M}_{\odot}$. The simulated galaxies lie in the stellar mass range between 2.75 and 10.97 $\times 10^{10 }\solarmass$. They are also star-forming at z=0 and exhibit a typical late-type disk galaxy component, although three of them lack an extended disk, showing a more spheroidal morphology. These simulated galaxies follow the general observational trends for MW-mass galaxies, and display a wide variety of properties, primarily due to the diversity in their merger histories \citep{grand2017}.
The Auriga simulations have been extensively used in recent years because they have demonstrated the ability of the galaxy formation model to replicate late-type galaxies with debris from merging activities \citep[for example in reproducing a Gaia Enceladus-like stream or linking the brightest stream to properties of satellites progenitors,][]{fattahi2019,vera2022}. These simulations have also been analyzed to study the properties of the stellar halos of galaxies, revealing diversity in their masses, density profiles, metallicities, shapes and ages, reflecting the stochasticity in their accretion and merger histories \citep{monachesi2016b,monachesi2019}. These properties have been compared to nearby MW-like galaxies, primarily from the GHOSTS Survey \citep{monachesi2016a,harmsen2017}, and have been found to be in good agreement with observations.
We then use this set of simulations, capable of reproducing many observable properties of MW-like galaxies, to analyze their kinematics  at distances representing the outer disk and the halo fields of NGC4945.

The TNG simulations are also a set of cosmological magneto-hydrodynamic simulations depicting the formation and evolution of galaxies. These simulations were run with three different physical box sizes, corresponding to cube volumes of approximately 50, 100, and 300 Mpc side lengths, referred to as TNG50, TNG100, and TNG300, respectively. The resolution increases with decreasing volume, with TNG50 \citep{nelson2019,pillepich2019} achieving a baryonic mass resolution of $8.5 \times 10^4 \textup{M}_{\odot}$. In this study, we specifically utilize MW and M31 analogues from TNG50 \citep{pillepich2023}. These analogues represent simulated galaxies with a stellar mass range between $10^{10.5}$ and $10^{11.2} \textup{M}_{\odot}$, exhibiting a disk-like morphology and visually identified spiral arms. Additionally, selected analogues meet specific criteria, including having no other galaxies within 500 kpc with a stellar mass greater than $10^{10.5}\textup{M}_{\odot}$ and a total mass of the host halo smaller than that typical of massive groups ($\leq 10^{13}\textup{M}_{\odot}$). These criteria yield a sample of 180 MW and M31 analogues for our analysis. 

\camila{Since the simulated galaxies have different disk sizes, we normalize their sizes by their optical radius (\ropt) to make a fair comparison with our measurements, although we bare in mind that it is unclear if stellar halos should scale with the \ropt of the disk. The \ropt of NGC 4945 is 12 kpc; this places our MUSE outer disk field at a distance of $\sim1.02 \times$\ropt and the halo field at $\sim2.9\times$\ropt along the major axis.  We rotate the simulated galaxies and select stellar particles from 0.95 to 1.1 \ropt and from 2.8 to 3.2 \ropt along the major axis, and within 1(2) kpc from the disk plane for the outer disk(halo) box in each simulated galaxy. These two selection criterion mimic the position of the MUSE outer disk and halo fields of NGC 4945, and we use the stellar particles inside those selected boxes to obtain the velocity measurements. Additionally, for comparison purposes, we selected the stellar particles in boxes at fixed distances of 12.2 kpc and 35 kpc, that mimic the physical locations of our MUSE fields. The results from this selection boxes are shown in Sec.~\ref{ap:sim_original}.}

For the two sets of simulated galaxies, we calculate the median LOS velocities for all the stellar particles and for only the accreted particles within each box of every galaxy. The top(bottom) panel of Fig.~\ref{fig:vel_simulation_ropt} shows the LOS velocity distribution of the box representing the halo(outer disk) region on the major axis, considering all the stellar particles (solid lines) and just the accreted ones (dashed lines), in orange for the Auriga galaxies and blue for TNG50.
We also mark with a black vertical line the LOS velocity, corrected for the systemic velocity of NGC~4945, of $-$44(110) \km obtained in this work for the halo(outer disk) of NGC~4945.
The grey area around black lines represents the uncertainty in our measurement.
We mark in green and pink the rotational halo and outer disk velocity of the MW and M31  \camila{at the corresponding distances of 1.02\ropt and 2.9\ropt. For the MW the \ropt is 12 kpc \citep{pilyugin2023} and for M31, it is 21.6 kpc \citep[transformed from an optical radius of 95.3';][]{devaucouleurs1991}. We marked the values a the corresponding $\sim$1.02\ropt and $\sim$2.9\ropt of both galaxies:} 200 and 15 \km for the MW 
\citep{bland-harwthorn2016,deason2017}{}{} and 210 and 200 for M31 \citep{zhang2024}.

The median LOS velocities of the Auriga galaxies in the halo selected box (top panel of Fig.~\ref{fig:vel_simulation_ropt}) are distributed between $-40$ and $\sim$160 \km, with a dominant peak at $\sim$20 \km. 
We virtually found no big differences between the total and accreted component distribution, which reflects that at those distances the Auriga simulations predict mostly an accreted component. For TNG50, the bulk of the total median LOS velocity distribution is around $\sim$30 \km, and there is, like in Auriga, no significant difference between the total and accreted component distribution. However, their values are more widely distributed than in Auriga, between \camila{$\sim-100$} and 240 \km. 

Our measured halo LOS velocity of NGC 4945 at \camila{$\sim$35} kpc along the major axis falls within the simulated values, in line with the kinematics of an accreted component of the stellar halo at that distance. The counter-rotation in our MUSE halo field indicates that at this distance of $\sim$35 kpc from the center of NGC~4945 the in-situ component is no dominant, assuming that this component should rotate with disk-like kinematics\footnote{Although we note that in few of the TNG galaxies some of the in-situ populations at $\sim$35 kpc have no rotation.}. 

For the box representing the outer disk field (bottom panel of Fig.~\ref{fig:vel_simulation_ropt}), both the Auriga and TNG50 galaxy samples show a peak in the LOS velocity distribution at around 150 \km when considering all particles. This clearly indicates the presence of a rotating disk at those distances. \camila{We highlight that, since NGC 4945 is edge-on, its LOS velocity in the outer disk field along the major axis can be regarded as a proxy of its rotational velocity.} 
We also show that a significant number of galaxies from both the TNG and Auriga simulations showcase elevated LOS velocities when solely considering the accreted particles. At this distance, the in-situ component is expected to dominate, so this feature can be attributed to a massive satellite that perturbed the disk in such a way that now both satellite and disks are aligned and rotating similarly \citep{Gomez2017}.

Our measured LOS velocity for NGC 4945 at \camila{$\sim$12.2} kpc along its major axis is 673 \km, and when corrected by its systemic velocity we obtained value of 110 \km, which is a good proxy of the rotational velocity of the galaxy, given its edge-on configuration.
This value falls within the distribution of LOS velocities obtained from the simulations.
Although our measured velocity of 673 \km  is slightly lower than the one measured in the disk using HI gas in a nearby position, this could be due to a potential contribution from the accreted halo component at this distance. We measured that the accreted halo component, at \camila{$\sim$35} kpc along the major axis, has a velocity of 519 \km, a bit lower but approximately consistent with the systemic velocity of the galaxy. This component might also be present at 12 kpc in the outer disk field, albeit much subdominant compared to the in-situ component, influencing the obtained LOS velocity.
Regardless, we conclude that the in-situ component is dominating at this distance.

Importantly, it is worth noting that the kinematic information of these two fields in NGC~4945 was crucial to conclude that at \camila{$\sim$35} kpc the accreted component of the stellar halo dominates, but at \camila{$\sim$12} kpc the in-situ component still dominates.

\begin{figure}
    \centering
    \includegraphics[width=\columnwidth]{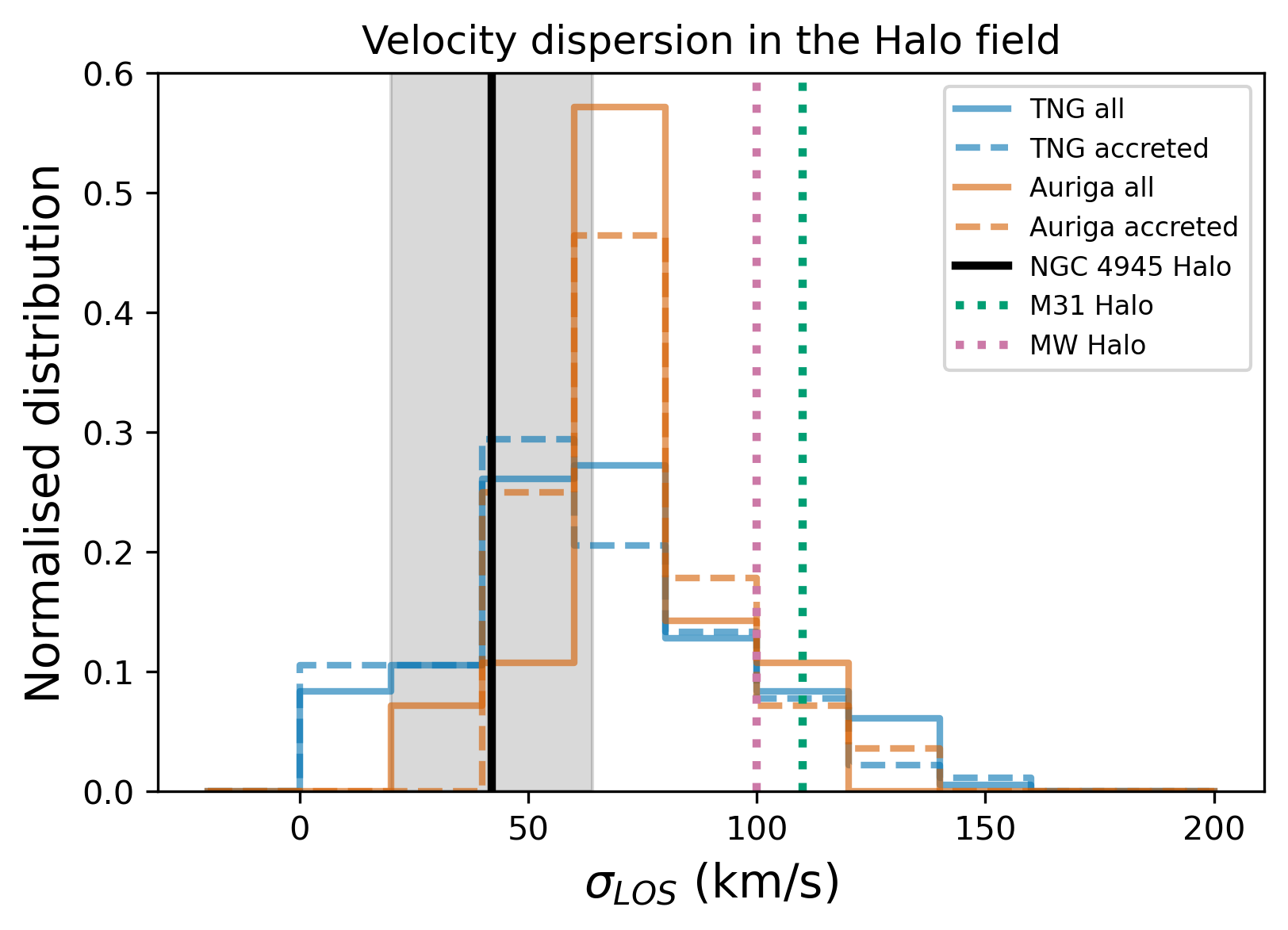}
    \includegraphics[width=\columnwidth]{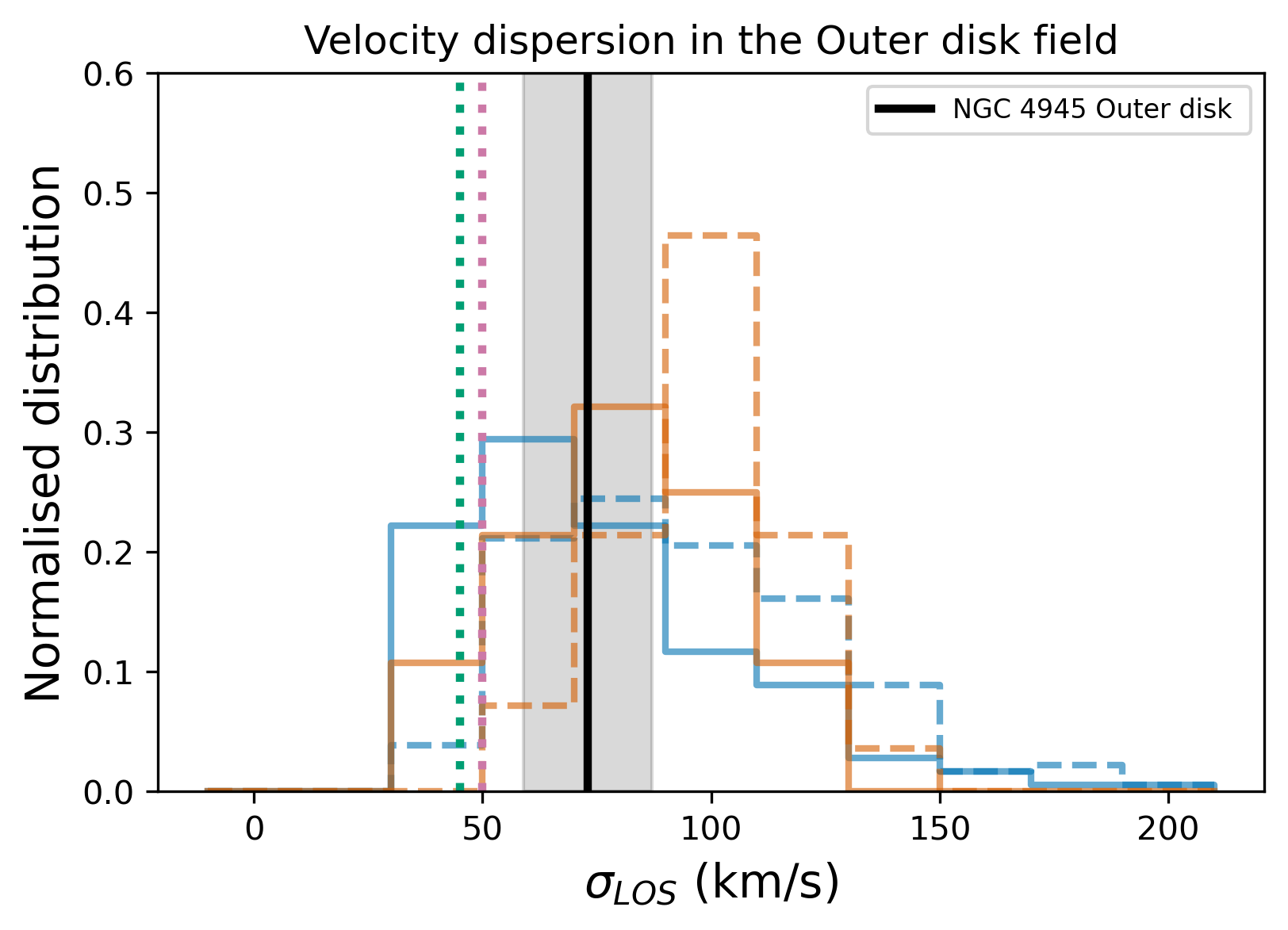}
    \caption{Velocity dispersion in the  LOS at galactocentric distances of \camila{1.02\ropt} (Upper) and \camila{2.9\ropt} (bottom) kpc along the major axis for MW/M31 like galaxies in the TNG50 (blue) and Auriga (orange) simulations. Black vertical lines are our velocity dispersion measurements for NGC 4945 at those distances, corrected for its systemic velocity. Green and pink dotted vertical lines show the halo velocity dispersion of the MW \citep[][]{bland-harwthorn2016} and M31 \citep{zhang2024}. Our measurements of NGC 4945 are in agreement with both set of cosmological simulations.}
    \label{fig:vel_disp_simulation}
\end{figure}

Fig.~\ref{fig:vel_disp_simulation} illustrates the distribution of velocity dispersion in the same regions, the halo and outer disk, used for the LOS velocity measurements in both Auriga and TNG simulations (Fig.~\ref{fig:vel_simulation_ropt}). Additionally, our velocity dispersion measurements for the halo and outer disk fields (42$\pm$22 \km and 73$\pm$14 \km, respectively) are indicated by black vertical lines. Furthermore, we include the velocity dispersion values for the MW \citep[50 and 100 \km for the outer disk and halo, respectively;][]{bland-harwthorn2016} and for M31 \citep[$\sim$45 and $\sim$110 \km for the outer disk and halo, respectively;][]{zhang2024}.

At a distance of \camila{$\sim$35} kpc, both simulations exhibit a peak around $\sim$70 \km in velocity dispersion, with values ranging from 0 to 140 \km for the Auriga simulations and 0 to 160 \km for the TNG simulations. Our measured LOS velocity dispersion of 42$\pm$22 \km is lower compared to the MW and M31 halo velocity dispersions. However, it falls well with the range represented by the cosmological simulations, albeit in the lower tail of values. Even considering the uncertainty, our measurement falls within the range of velocity dispersion values depicted by the simulations. This value also supports our scenario of an accreted halo, with a low counter-rotation, and this lower velocity dispersion may indicate that we are capturing some colder halo structure.

At a distance of \camila{$\sim$12.2} kpc, the peak velocity dispersion ranges between 50 and 80 \km in both sets of simulations, with values ranging from \camila{30} to \camila{190} \km in the TNG simulations and between 30 and \camila{150} \km in the Auriga simulations. Our measured velocity dispersion in the outer disk field is 73$\pm$14 \km, falling squarely within the range exhibited by the simulations. It is important to note that the velocity dispersion observed in this field exceeds that of the halo and exceeds what is typically expected in a disk. The positioning of the outer disk field, particularly with one end closer to the disk, where we observe a drastic increase in stellar density in that corner, may account for this higher velocity dispersion. We attribute this elevated dispersion to a combination of contributions from both ordered motion in the disk and stars with kinematics more akin to the halo, resulting in an overall increase in velocity dispersion.

\section{Summary and Conclusions}

In this study we present the first stellar kinematical measurements for a diffuse stellar halo of a MW-like galaxy beyond the LG, NGC 4945. NGC 4945, at a distance of 3.56 Mpc, is an ideal target to study since it features a typical stellar halo for a MW-mass galaxy, in terms of mass and metallicity. Thus, the results from this study are very useful into understanding the stellar halo formation and origin of MW-mass galaxies. 
We use new deep MUSE observations to measure the mean heliocentric LOS velocities and velocity dispersions in two fields along the major axis of the nearby galaxy NGC 4945. One of these fields is located at \camila{$\sim$35} kpc from the center of NGC 4945, representing the halo field, while the second one is situated at \camila{$\sim$12} kpc, characterizing the outer disk field. These fields were strategically chosen along the major axis in order to disentangle the origin of its stars, as models predict that the in-situ halo population is more noticeable along the major axis. The distance of the halo field was chosen so as to reach as far out as possible into the stellar halo but at the same time to have enough RGB spectra to stack be able to reach the required S/N to do kinematical measurements. The outer disk field serve as the control disk field for velocity comparisons. NGC4945 is an edge-on galaxy, thus a velocity measurement along its major axis is a good proxy of its rotational velocity. 

By combining the MUSE data with existing HST catalogs at those locations, we were able to extract the spectra of individual AGB and RGB stars, in each of these fields, down to 2 magnitudes below the TRGB. 

Only for the brightest stars in the halo field, with F814W<23.72, the S/N values of their spectra were greater than 4, allowing us to make kinematical measurements of individual stars and discriminate between foreground MW stars and potential stars belonging to NGC 4945, brighter than the TRGB. For the remaining RGB stars in the halo field and the AGB and RGB stars in the outer disk we co-added their individual spectra to obtain a higher S/N stacked spectrum for each field.  We summarize our main findings in the following: 
\begin{itemize}

   \item For the halo field, we co-added 53 RGB stars reaching a S/N= 9.4 for its stacked spectrum.  We measured a mean heliocentric LOS velocity of 519$\pm$12\km and a velocity dispersion of 42$\pm$22 \km. Given NGC 4945 systemic velocity of 563 \km, this measurement shows that the halo of NGC 4945 at \camila{$\sim$35} kpc from its center along the major axis is slightly counter-rotating, with a rotational velocity of $-44$ \km, which demonstrates its accreted origin. \\

   \item Most of the stars brighter than the TRGB in the halo field are MW foreground stars, based on their velocities measured. However, we found 3 stars in the halo field, brighter than the TRGB, with velocities larger than 400 km/s. These stars are strong candidates to belong to NGC 4945 stellar halo based on their velocities only. According to their position in the CMD, these stars would be one AGB, one BHeB and one RHeB. \\
     
   \item For the outer disk field, we co-added 1122 RGB stars along with 70 AGB stars, resulting in a stacked spectrum with S/N= 16.7 and a mean heliocentric LOS velocity of 673$\pm$11 \km and a velocity dispersion of 73$\pm$14 \km. This is consistent with the mean HI velocity of the disk at a nearby position \citep[$\sim$700 \km,][]{koribalski2018}.\\

   \item We compare our results with two sets of cosmological simulations of MW-like galaxies, Auriga and TNG50, where we measure the LOS kinematics of the simulated galaxies in two boxes, at \camila{1.02\ropt} and \camila{2.9\ropt} kpc along the major axis of the galaxies placed edge-on, representing the outer disk and halo fields. We find that our measurements are in agreement with the simulated values and are consistent with an accreted halo at \camila{$\sim$35} kpc and in-situ disk kinematics at \camila{$\sim$12.2} kpc.
   
\end{itemize}
 
The resolved stellar halo velocity measurement of NGC 4945 with MUSE sets a new standard for studying stellar halos in galaxies beyond the Local Group. The results presented in this study highlight the existence of an accreted flattened stellar halo with a slight counter-rotation. Importantly, this work adds a third galaxy with these types of measurements --- the only other two galaxies so far for which their diffuse stellar halos has been studied kinematically from its resolved stars are the MW and M31.

The findings of this research pave the way for a more comprehensive exploration of the kinematics in distant galaxies, offering unprecedented opportunities to unravel the mysteries of stellar halo formation and evolution.
The successful application of this methodology and analysis demonstrates the potential for future studies using next-generation telescopes like the ELT. 
With its 39 meter diameter segmented mirror, this telescope will facilitate the resolution of individual halo stars not only for nearby but also for more distant galaxies compared to what is currently available with instruments like MUSE. The MOSAIC instrument is a multi-object spectrograph with a large FoV,  40 times larger than that of MUSE, which will be exquisite to kinematically map larger regions of the halos in nearby galaxies. MOSAIC will have a visible mode which will enable observations of nearly 200 objects simultaneously, with both medium and high spectral resolving power (R $\sim$ 5000 and R $\sim$ 20000). In addition, its NIR mode will observe 80 objects simultaneously, as well as dedicated sky fibers to account for the strong background in the NIR regime. To have dedicated sky fibers will improve significantly the subtraction of background in the target objects. Additionally, the HARMONI instrument will be a visible and near-infrared integral field spectrometer, similar to MUSE,  with a smaller field of view, but higher resolution. Thus HARMONI will be more suitable to study regions of the halos in more distant galaxies, offering a potential to achieve detailed results with comparable or even shorter integration times than those employed in this study. These two instruments will enable detailed studies of kinematics and metallicity for individual stars in more nearby and in distant galaxies. This will allow us to apply our technique to more galaxies in the future, expanding significantly the statistical sample to better understand the origin of stellar halos.

\begin{acknowledgements}
\camila{We thank the anonymous referee for useful comments and suggestions that helped to improve the quality of this paper. We also wish to thank Arianna Dolfi for providing the optical radius for the TNG50 galaxies. }
C.B. acknowledges the support provided by ANID/"Beca Doctorado Nacional" 21211381.
A.M. and C.B. gratefully acknowledge support by the ANID BASAL project FB210003 and FONDECYT Regular grant 1212046. A.M. and FAG acknowledge funding from the Max Planck Society through a “PartnerGroup” grant. FAG acknowledges support from ANID FONDECYT Regular 1211370 and ANID Basal Project FB210003. EFB acknowledges support from the National Science Foundation through grant NSF-AST 2007065. \camila{We acknowledge the usage of HyperLEDA database (http://leda.univ-lyon1.fr/).}

\end{acknowledgements}

\bibliographystyle{aa}
\bibliography{ref} 

\appendix

\section{Velocity dispersion tests}
\label{ap:sigma_test}
\begin{figure}
    \centering
    \includegraphics[width=\columnwidth]{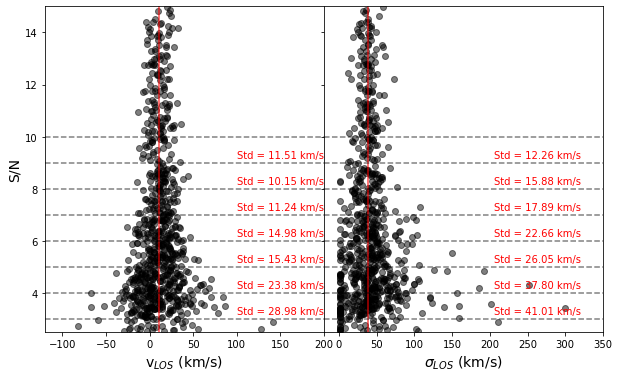}
    \caption{\camila{Scatter in the measurements of the velocity and velocity dispersion for stacked spectra with a dispersion of 40 \km, resembling spectra of stacked RGB stars, as a function of the S/N. 
    The red vertical lines indicate the median values of velocity and velocity dispersion, 11 and 39 \km respectively. We also report the standard deviation at various levels of S/N, which decreases as the S/N increases.}}
    \label{fig:sigma_width40}
\end{figure}

\camila{Motivated by the fact that our measured velocity dispersion in the halo field falls below the MUSE resolution of $\sim$50 \km, we  repeated the test shown in Fig.~\ref{fig:sigma_test}. To do this we repeated the methodology outlined in Sec.~\ref{sec:ppxf} for a stack spectrum, but instead of using a Gaussian distribution with a width of 100 \km, we used a velocity distribution with a width of 40 \km, mimicking the velocity dispersion we obtained in the halo field. 
As shown in Fig.~\ref{fig:sigma_width40}, this test confirms the pPXF's capability to recover such low velocity dispersions, particularly for spectra with S/N$>$8. For stacked spectra with S/N$\leq8$, the possibility increases that pPXF may either fail to measure the velocity dispersion (resulting in a value of 0) or mismeasure it significantly (providing values much higher than 40 \km).
Unfortunately, due to the limitations of MUSE's resolution, we cannot resolve velocity dispersions lower than this value. Therefore, as mentioned in Sec.~\ref{sec:halo_stacked}, we can only confirm that the velocity dispersion in our halo field is lower than 50\km.}

\section{Comparing LOS velocities using fixed boxes at 12.2 and 35 kpc}
\label{ap:sim_original}
\begin{figure}
    \centering
    \includegraphics[width=\columnwidth]{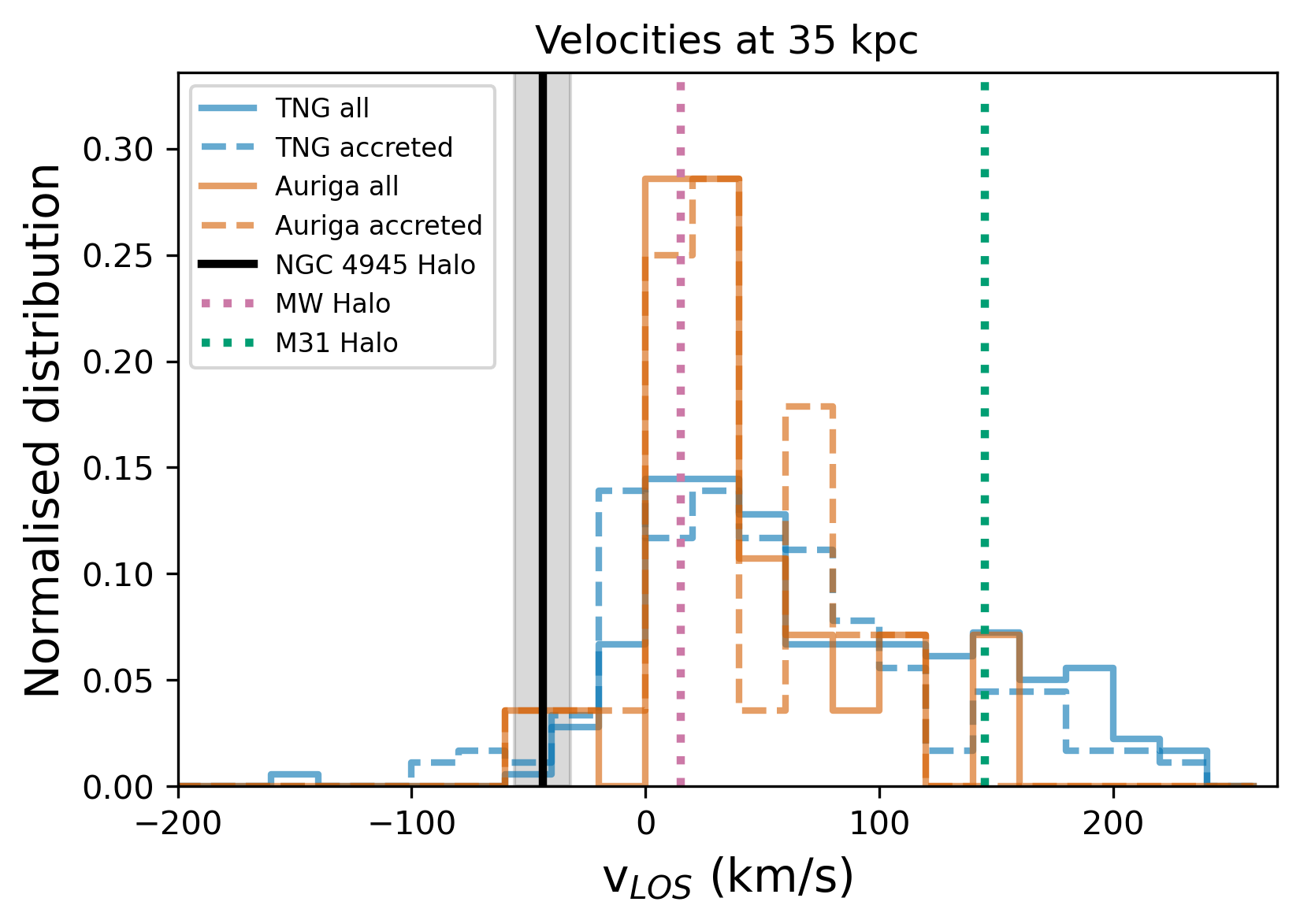}
    \includegraphics[width=\columnwidth]{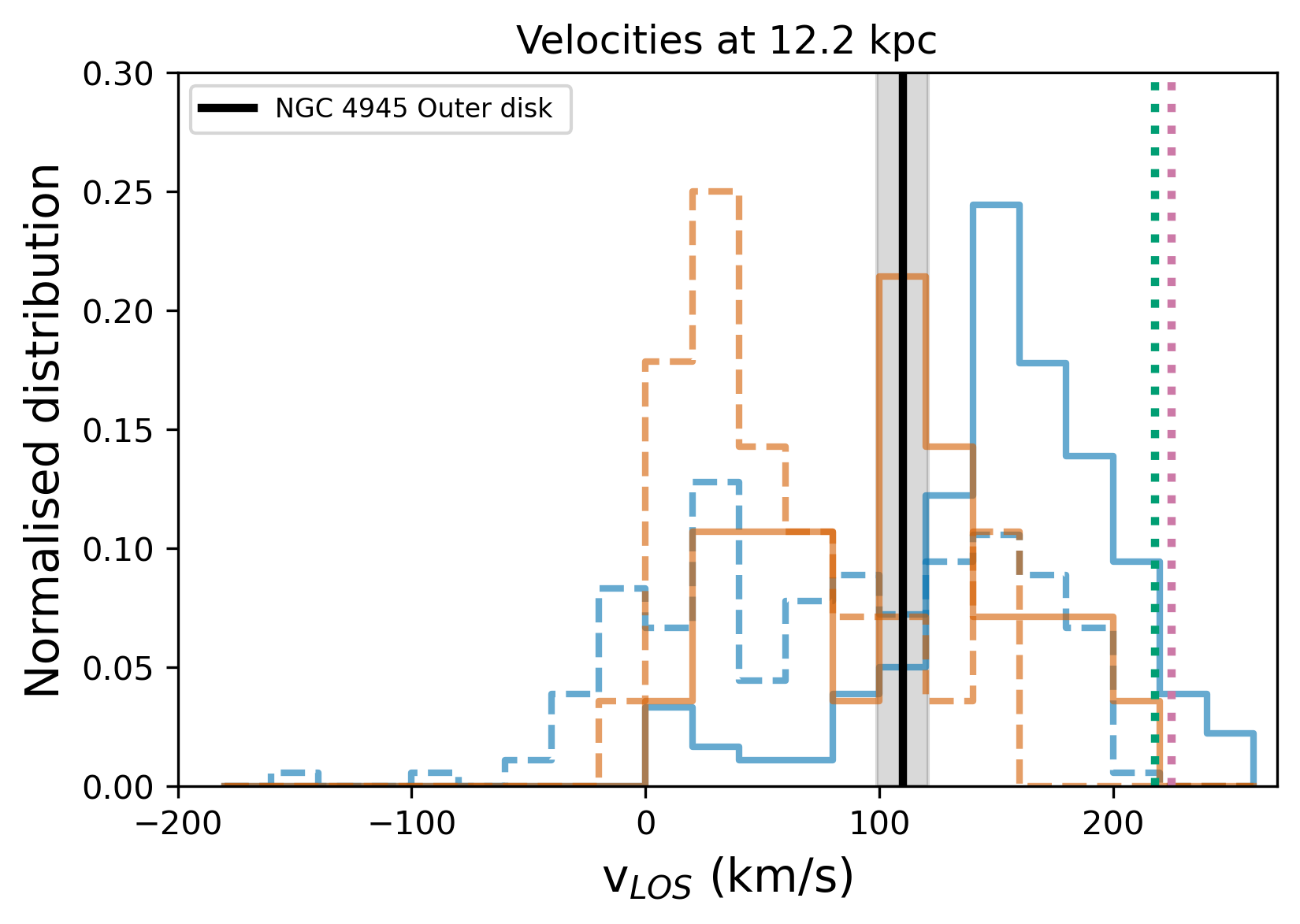}
    \caption{Median  LOS  velocity distribution at galactocentric distances of $\sim$35 (Upper) and 12.2 (bottom) kpc along the major axis for MW/M31 like galaxies in an edge-on configuration, in the TNG50 (blue) and Auriga (orange) simulations. Black vertical lines are our velocity measurements for NGC 4945 at those distances, corrected for its systemic velocity. Green and pink dotted vertical lines show the halo rotational velocity of the MW \citep[halo and outer disk,][respectively]{deason2017, bland-harwthorn2016} and M31 \citep[halo and outer disk,][respectively]{ibata2005,zhang2024}. Our velocity measurements of the stellar halo of NGC~4945 at $\sim$35 kpc are in agreement with those of the accreted component of the simulated galaxies at $\sim$35 kpc. At a distance of 12.2 kpc simulations show that our results are consistent with the disk in-situ component dominating the field, with little contribution from the accreted component.}
    \label{fig:apx_vel_simulation}
\end{figure}

\begin{figure}
    \centering
    \includegraphics[width=\columnwidth]{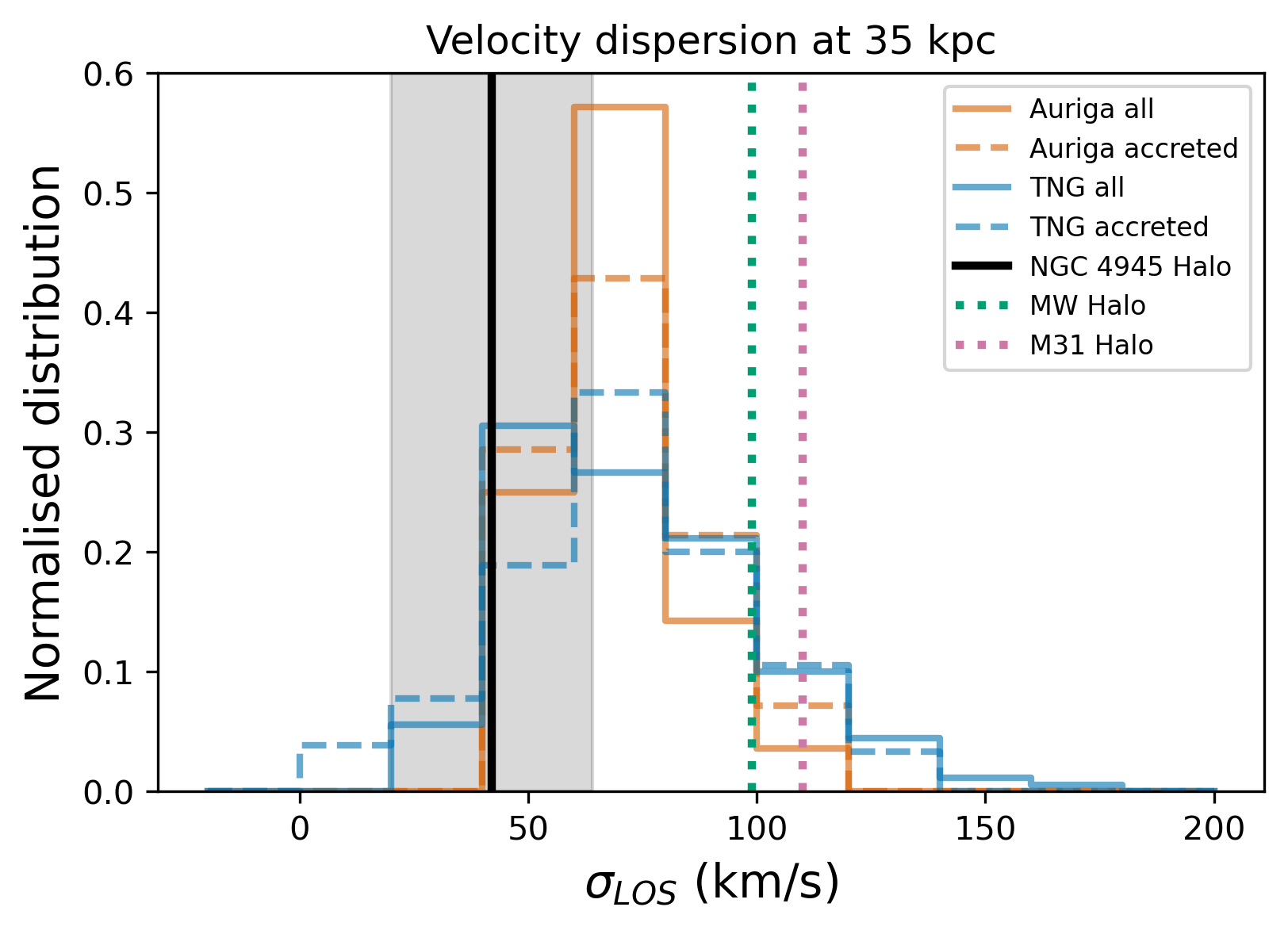}
    \includegraphics[width=\columnwidth]{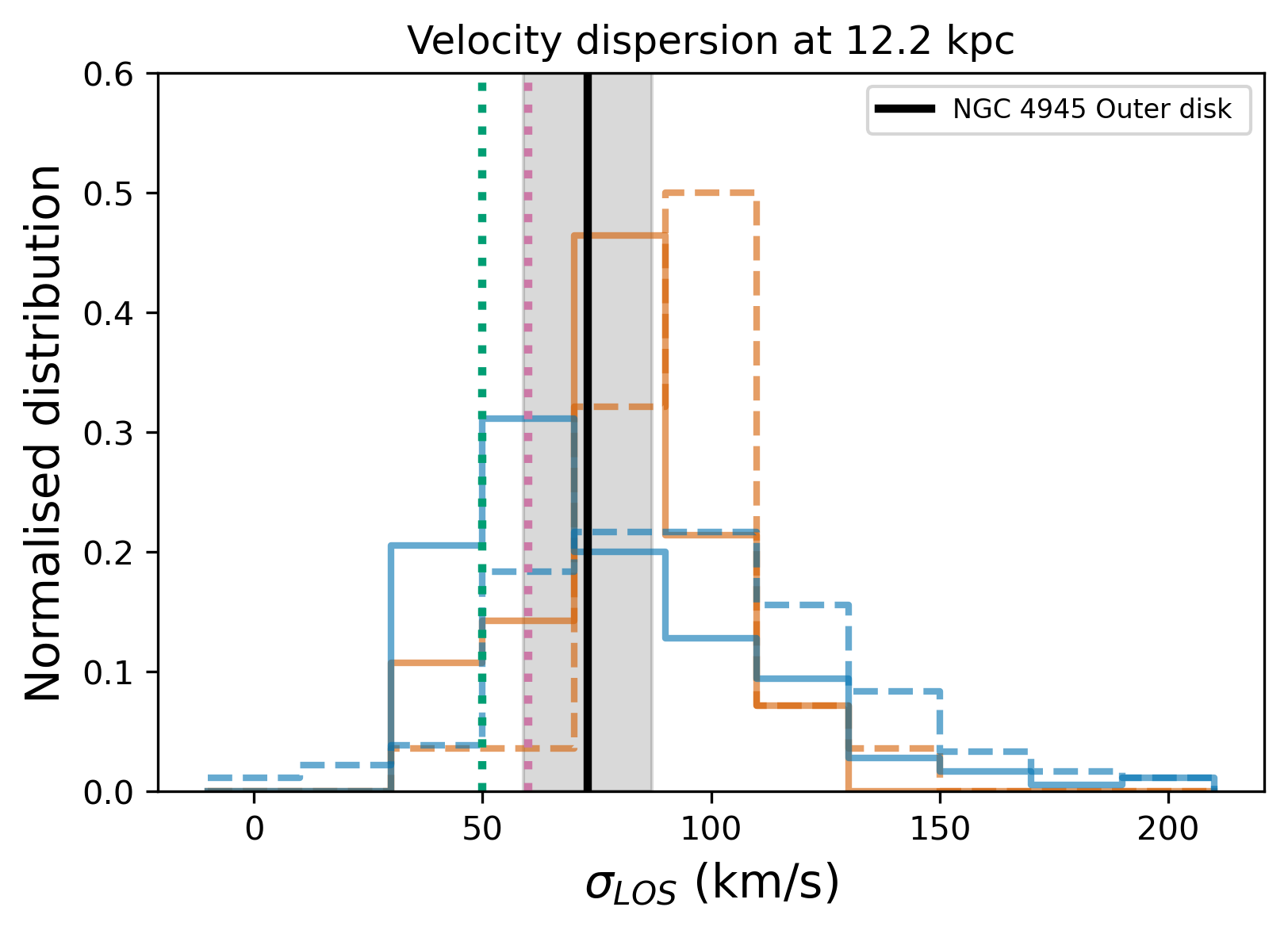}
    \caption{Velocity dispersion in the  LOS at galactocentric distances of $\sim$35 (Upper) and 12.2 (bottom) kpc along the major axis for MW/M31 like galaxies in the TNG50 (blue) and Auriga (orange) simulations. Black vertical lines are our velocity dispersion measurements for NGC 4945 at those distances, corrected for its systemic velocity. Green and pink dotted vertical lines show the halo velocity dispersion of the MW \citep[][]{bland-harwthorn2016} and M31 \citep{zhang2024}. Our measurements of NGC 4945 are in agreement with both set of cosmological simulations.}
    \label{fig:apx_vel_disp_simulation}
\end{figure}

\camila{For comparison purposes, we plotted the LOS velocity and velocity dispersion at fixed distances of 12.2 and $\sim$35 kpc from the center of each galaxy along the major axis (Fig.~\ref{fig:apx_vel_simulation} and \ref{fig:apx_vel_disp_simulation}).}

\camila{At $\sim$35 kpc, both sets of simulations do not show significant differences compared to the results obtained by normalizing the distance to 2.9\ropt. Similarly, there are no significant differences when we fix the boxes at 12.2 kpc, compared to the results using a normalized distance of 1.02\ropt.}


\end{document}